\documentclass[universe,review,submit,pdftex,moreauthors]{Definitions/mdpi} 
\firstpage{1} 
\pubvolume{1}
\issuenum{1}
\articlenumber{0}
\pubyear{2025}
\copyrightyear{2025}
\datereceived{ } 
\daterevised{ } 
\dateaccepted{ } 
\datepublished{ } 
\hreflink{https://doi.org/} 

\usepackage{longtable}

\Title{On the X-ray Emission From Supernovae, and Implications for the Mass-Loss Rates of their Progenitor Stars.}
\TitleCitation{X-ray Emission from Supernovae}

\Author{Vikram Dwarkadas $^{1,\dagger}$\orcidA{},}

\AuthorNames{Vikram Dwarkadas}

{
        \AuthorCitation{Dwarkadas, V}
        }

\address{%
$^{1}$ \quad vikramvd@uchicago.edu}

\corres{Correspondence: vikramvd@uchicago.edu; Tel.: +01-773-834-4668}

\firstnote{Current address: Dept.~of Astronomy and astrophysics, University of Chicago, 5640 S Ellis Ave., ERC 569, Chicago, IL 60637}  

\abstract{We summarize the X-ray emission from young SNe. Having accumulated data on most observed X-ray SNe, we display the X-ray lightcurves of young SNe. We also explore the X-ray spectra of various SN types. The X-ray emission from Type Ib/c SNe is non-thermal. It is also likely that the emission from Type IIP SNe with low mass-loss rates (around 10$^{-7} \, \msun \,$ yr$^{-1}$) is non-thermal. As the mass-loss rate increases, thermal emission begins to dominate. Type IIn SNe have the highest X-ray luminosities, and are clearly thermal. We do not find evidence of non-thermal emission from Type IIb SNe. The aggregated data are used to obtain approximate mass-loss rates of the progenitor stars of these SNe. Type IIP's have progenitors with mass-loss rates $< 10^{-5}\, \msun \,$ yr$^{-1}$, while Type IIn progenitors generally have mass-loss rates $> 10^{-3}\, \msun $ yr$^{-1}$. However, we emphasize that the density of the ambient medium is the important parameter, and if it is due to a non-steady outflow solution, it can not be translated into a mass-loss rate. }


\newcommand\msun{\ensuremath{\rm M_{\odot}}}

\newcommand{\beqn}{\begin{eqnarray}}
\newcommand{\eeqn}{\end{eqnarray}}
\newcommand{\be}{\begin{equation}}
\newcommand{\ee}{\end{equation}}
\newcommand{\noi}{\noindent}

\newcommand{\mdot}{\mbox{$\dot{M}$}}

\newcommand{\bfig}{\begin{figure}[H]}
\newcommand{\efig}{\end{figure}}

\begin{document}
\nolinenumbers



\section{Introduction:} Core-collapse Supernovae (SNe) arise from massive stars. These stars have strong mass-loss in the form of stellar winds, with mass-loss rates generally ranging from 10$^{-7}$ to 10$^{-4}$ $\msun$ yr$^{-1}$, and therefore lose a considerable amount of mass over their evolution. When the star explodes as a SN, the resulting shock wave expands in this wind-blown medium. The evolution of the SN shock wave, and the resulting emission, depends on the structure and density profile of the medium. The rapidly expanding high-velocity shock wave heats up the ambient medium to high temperatures, leading to copious amounts of thermal X-ray emission via thermal bremsstrahlung combined with line emission \cite{chevalier82b}. The shocks also accelerate particles, both electrons and protons, to high energies. The accelerated electrons can give rise to nonthermal X-ray emission, via synchrotron or inverse Compton processes \cite{cf06,cfn06,cf17}.  X-ray emission is thus one of the signposts of the shock wave interaction with the circumstellar medium \cite{cf03}.

The radiation signatures from the expansion of the SN shock wave can be used to trace the density profile of the medium in which the SN is expanding. In core-collapse SNe, this medium is formed by mass-loss from the progenitor star. Thus the emission signatures can inform us of the mass-loss evolution from the pre-SN star, especially in the final stages of SN evolution. Wind mass-loss rates in the final stages of a massive star's evolution are otherwise notoriously difficult to obtain.

The first SN to be discovered in X-rays was SN 1980K \cite{canizaresetal82}. X-rays were detected with the Imaging Proportional Counter aboard the Einstein satellite 35 days after maximum light. A second observation 82 days after maximum also detected X-ray emission, giving birth to the era of X-ray SNe. The pace of discovery was slow in the initial stages. By 1995 only 5 X-ray SNe were known \cite{schlegel95}. The number of X-ray SNe increased dramatically with the launch of Chandra and XMM-Newton satellites in 1999, followed by Swift in 2004. By 2003, \cite{il03} had listed 15 X-ray supernovae. By 2006, \cite{schlegel06} had increased this number to 25. \cite{immler07} noted, on the occasion of the 20th anniversary of SN 1987A,  that 32 SNe had been discovered in X-rays, all of which were core-collapse SNe. \cite{dg12} published a paper showing the X-ray lightcurves of most SNe that had been detected to date, and found that 42 X-ray SNe had been detected. {Subsequent} papers continued to increase this number \cite{dwarkadas14, drrb16}. Discovery of the first Type Ia SN in X-rays was published in 2018 by \cite{bocheneketal18}. SN 2012ca was a Type Ia-CSM, a Type Ia which shows an H$\alpha$ line in the optical spectrum, at least for a short time. {The classification of SN 2012ca as a Type Ia was disputed by \cite{inserraetal14}, who labelled it as a stripped envelope core-collapse SN. However,  \cite{foxetal15} have argued that the spectra are more consistent with a thermonuclear explosion.}  To date, although more than a 100 SNe have been observed in X-rays, SN 2012ca remains the only non-core-collapse SNe to be detected in X-rays. The remaining X-ray SNe arise from massive star progenitors, although it has been postulated that Ca-rich SNe may arise from low mass stars. A fair number, especially the Type Ib/c SNe, are {theorized to have binary progenitors \cite{lymanetal16,taddiaetal18,prenticeetal19,barbarinoetal21,solaretal24}, though not all Ib/c SNe arise from a binary population \cite{karamehmetogluetal23}.}

Over the last 30 years, a few reviews of the X-ray emission from SNe have been published. Theoretical aspects of the X-ray emission have been comprehensively dealt with by \cite{cf03}, and later expanded on in \cite{cf17}. Observationally, many authors have tried to categorize the emission from SNe since the early review by Schlegel \cite{schlegel95}, as listed above. Two recent reviews \cite{chandra17,chandra18} concentrated mainly on SNe in dense environments, especially the class of Type IIn SNe.

In this paper we review the X-ray emission from SNe, and present the X-ray lightcurves, or upper limits, for over a 100 SNe that have been observed in X-rays. We further study the spectra of a sample of SNe of various types, to gain an insight into their emission mechanism. We then use this aggregation of SNe, combined with knowledge of their spectra, to explore the mass-loss rates of massive stars that collapse to form SNe. While we have tried to include as many SNe as we can find, we do not claim that our compilation is complete. We ask readers who are aware of any SNe missing from Table~\ref{tab:xraysne} to kindly inform the author, so that we may include them in subsequent articles and plots.
 
The remainder of this paper proceeds as follows. In \S2 we present the lightcurves of all  X-ray {observed} SNe {known to the author}. We also plot them grouped by type to distinguish between the luminosities of various classes of SNe. In \S3 we study the X-ray spectra of various types of SNe. We discuss whether the spectra can {allow us to} distinguish between thermal or nonthermal emission. \S4 then uses the lightcurves, and the information gleaned from the spectra of various types, to study and discuss the mass-loss rates inferred for SNe of  different types, and the implications for their progenitors. Our conclusions are given in \S 5.

\section{X-ray Supernovae}

\begin{figure}[H]
\begin{center}
 \includegraphics[width=\columnwidth]{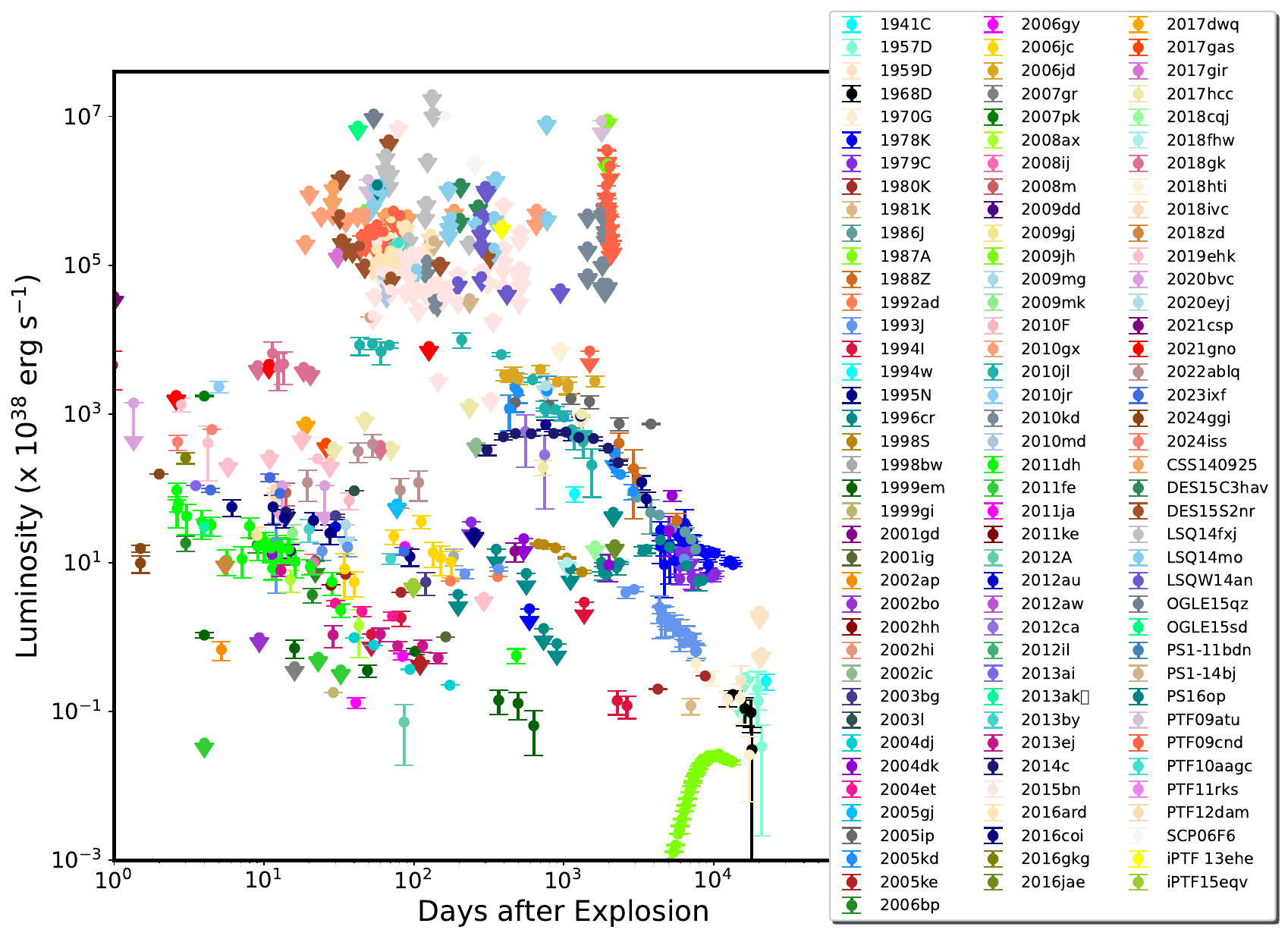} 
 \caption{X-Ray Lightcurves of most observed X-Ray SNe. SNe that were not detected but have upper limits are plotted with a downward facing arrow. There are currently 115 SNe on this list.
}
\label{fig:xraysne}
\end{center}
\end{figure}

Analysis of an X-ray observation of a SN provides a source flux. Combined with the distance to the SN (which may not always be well known), this provides a luminosity at a given epoch. Plotting the luminosity over a series of epochs gives rise to the X-ray lightcurve of the SN.  Over the years we have compiled a library of X-ray lightcurves of young SNe, both those that have appeared in the literature and those analysed by us. This process started with the publication of a first compilation of lightcurves in 2012 (\cite{dg12}, to which we have gradually added more over the years (\cite{dwarkadas14} and \cite{drrb16}) Many of these lightcurves are available in an online database that we have created for this purpose (\cite{rd17}, which can be viewed at {\tt kronos.uchicago.edu/snax}. The stability of the database was improved, and it was made accessible in 2020 \cite{ndr20}. 

Figure~\ref{fig:xraysne} displays the lightcurves of most X-ray SNe. The SNe themselves are listed in Table~\ref{tab:xraysne}. There are 115 SNe in this list, comprising 656 different data points. Note that each datapoint does not necessarily correspond to one observation, as some of them may be made from multiple observations that have been {combined} together. 

\begin{figure}[H]
\begin{center}
 \includegraphics[width=\columnwidth]{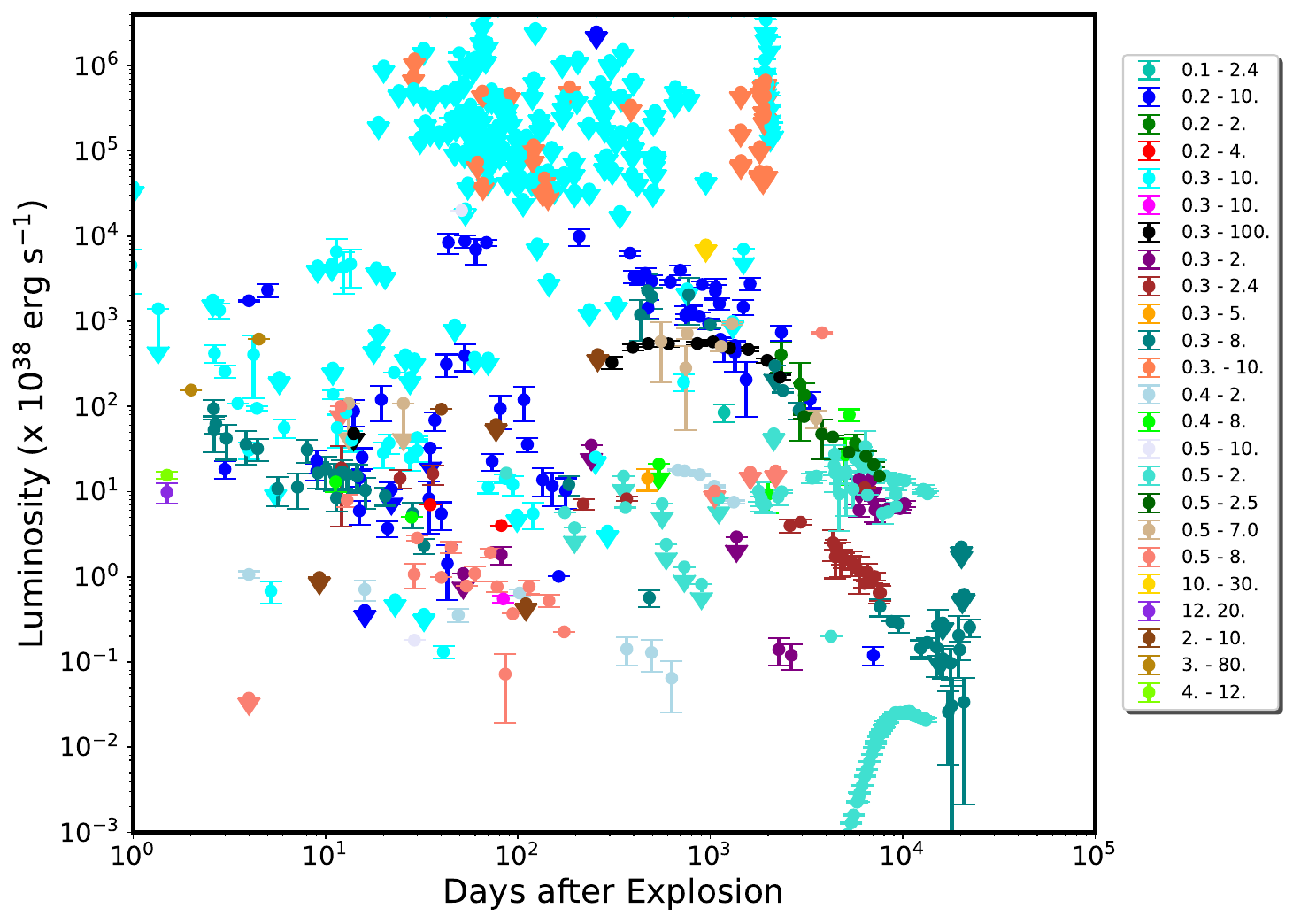} 
 \caption{The various bands in which X-ray SN luminosities are quoted in the literature.
}
\label{fig:bands}
\end{center}
\end{figure}

Unfortunately, fluxes are quoted by different authors in different bands, making comparison difficult. For each SN, we have made sure to list only the flux in a given band which has most of the data. For SNe observed with ROSAT this has generally meant 0.3-2.4 keV. For those observed later, authors list the result in a variety of bands, such as 0.3-8 keV, 0.2-10 keV, 0.3-10 keV, 0.5-8 keV,  0.5-10 keV, and so on. This often meant leaving out some data points not in the same band. Various energy ranges that have been used for deriving the X-ray luminosity are shown in Figure~\ref{fig:bands}. A useful common band for the Chandra, XMM-Newton and Swift telescopes, which detect the majority of the SNe, is 0.3-8 keV. We urge all authors to report fluxes in this band. The addition of NuStar can result in a much broader band, and some authors have used 0.3-100 keV. However, most SNe detected in the last two decades can and should be listed in the 0.3-8 keV band, making comparison easier.

 If SNe were observed but not detected, upper limits are shown, represented by downward facing arrows. Upper limits in X-ray astronomy are difficult to obtain accurately \cite{kashyapetal10}, with various techniques being used by different authors. Authors may also quote upper limits to 1, 2, or 3$\sigma$ accuracy, sometimes without specifying, thus making them somewhat difficult to compare. Upper limits from the literature are taken from the referenced papers, and the reader is directed to the appropriate paper to understand how the upper limits were obtained.

\begin{longtable}{l|l|l}
\caption[longtable]{List of all X-ray SNe plotted in Figure~\ref{fig:xraysne}, along with the type of SN and the citation of the paper from which the data were taken. \label{tab:xraysne}} \\
\hline
\textbf{SN Name}	& \textbf{Type}	& \textbf{References}\\
\hline
1941C		& II			& \cite{rd20} \\
1957D		& II			& \cite{rd20} \\
1959D		& IIL  		& \cite{rd20} \\
1968D		& II 			& \cite{rd20} \\
1970G 		& IIL			& \cite{rd20} \\
1978K.              & IIn			& \cite{schlegeletal04, zhaoetal17} \\
1979C		& IIL			& \cite{immleretal05} \\
1980K               & IIL. 		& \cite{fridrikssonetal08} \\
1981K		& II			& \cite{immleretal07a} \\
1986J		& IIn			& \cite{houck05} \\
1987A		& IIpec		& \cite{franketal16,ravietal24}$^1$ \\
1988Z		& IIn			& \cite{sp06} \\
1992ad		& IIP			& \cite{bregmanetal03} \\
1993J		& IIb			& \cite{chandraetal09b,dbbb14} \\
1994I		& Ic			& \cite{immleretal02} \\
1994W 		& II			& \cite{schlegel99} \\
1995N		& IIn			& \cite{chandraetal05,zampierietal05} \\
1996cr		& IIn			& \cite{quirolavasquezetal19} \\
1998bw             & Ic                 & \cite{kouveliotouetal04}\\
1998s.               & IIn			& \cite{pooleyetal02} \\
1999em             & IIP                 & \cite{pooleyetal02} \\
1999gi               & IIP                 & \cite{schlegel01} \\
2001gd              & IIb			& \cite{pereztorresetal05} \\
2001ig               & IIb			& \cite{sr02} \\
2002ap		 & IIb			& \cite{sutariaetal03} \\
2002bo		 & 1a			& \cite{hughesetal07} \\
2002hh		 & IIn			& \cite{pl02} \\
2002hi 		 & IIn			&\cite{pl03a} \\
2002ic		 & Ia-CSM		& \cite{hughesetal07} \\
2003L		 & 1c			& \cite{soderbergetal05} \\
2003bg		 & 1c			& \cite{soderbergetal06} \\
2004dj		 & IIP		& \cite{chakrabortietal12} \\
2004dk		 & 1b 		& \cite{pooleyetal19} \\
2004et		 & IIP		& \cite{misraetal07} \\
2005gj		 & 1a-CSM	& \cite{hughesetal07} \\
2005ip               & IIn                	& \cite{katsudaetal14} \\
2005kd              & IIn 		& \cite{drrb16} \\
2005ke		 & Ia			 & \cite{hughesetal07} \\
2006bp		 & IIP                & \cite{immleretal07b} \\
SCP06F6          & SLSN-I         & \cite{levanetal13} \\
2006gy		 & SLSN-II	& \cite{smithetal07}\\
2006jc 		& Ib			& \cite{immleretal08} \\
2006jd		& IIn			& \cite{chandraetal12} \\
2007gr		& Ic			& \cite{soderbergetal10} \\
2007pk		& IIn			& \cite{immleretal07c} \\
2008m		& II			& \cite{immler10} \\
2008ax		& IIb			& \cite{romingetal09} \\
2008ij		& IIP			& \cite{immleretal09a} \\
2009dd		& II			& \cite{immleretal09b} \\
2009gj 		& IIb			& \cite{immleretal09c} \\
2009mg		& IIb			& \cite{oatesetal12} \\
2009mk 		& IIb			& \cite{ri10} \\
PTF09atu		& SLSN-I		& \cite{marguttietal18} \\
PTF09cnd		& SLSN-I		& \cite{marguttietal18} \\
2009jh		& SLSN-I		& \cite{marguttietal18} \\
PTF10aagc	& SLSN-I		& \cite{marguttietal18} \\
2010F		& II			& \cite{rim10} \\
2010gx		& SLSN-I		& \cite{marguttietal18} \\
2010jl		& IIn			& \cite{chandraetal15} \\
2010jr		& II			& \cite{immleretal10} \\
2010kd		& SLSN-I		& \cite{marguttietal18} \\
2010md		& SLSN-I		& \cite{marguttietal18} \\
2011dh		& IIb			& \cite{soderbergetal12,maedaetal14} \\
2011ja		& IIP			& \cite{chakrabortietal13} \\
2011fe		& Ia			& \cite{marguttietal12} \\
2011ke		& SLSN-I		& \cite{marguttietal18} \\
PS1-11bdn	& SLSN-I		& \cite{marguttietal18} \\
PTF11rks		& SLSN-I	 	& \cite{marguttietal18} \\
2012a		& IIP			& \cite{pooley12} \\
2012au		& Ib			& \cite{kambleetal14} \\	
2012ca		& Ia-CSM		& \cite{bocheneketal18} \\
PTF12dam	& SLSN-I        &\cite{marguttietal18} \\
2012il		& SLSN-I		& \cite{marguttietal18} \\
2013ai		& II			& \cite{marguttietal13a} \\
2013ak		& IIb			& \cite{marguttietal13b} \\
2013by 		& IIL			& \cite{marguttietal13c,blacketal17} \\
2013ej		& IIP			& \cite{chakrabortietal16} \\
iPTF13ehe        & SLSN-I		& \cite{marguttietal18} \\
CSS140925	& SLSN-I		& \cite{marguttietal18} \\
2014c		& 1b/IIn		& \cite{thomasetal22,brethaueretal22} \\
LSQW14an	& SLSN-I		& \cite{marguttietal18} \\
LSQ14fxj		& SLSN-I		& \cite{marguttietal18} \\
LSQ14mo		& SLSN-I		& \cite{marguttietal18} \\
PS1-14bj		& SLSN-I		& \cite{marguttietal18} \\
DES15C3hav    & SLSN-I		& \cite{marguttietal18} \\
DES15S2nr	& SLSN-I		& \cite{marguttietal18} \\
OGLE15qz	& SLSN-I		& \cite{marguttietal18} \\
OGLE15sd	& SLSN-I		& \cite{marguttietal18} \\
2015bn		& SLSN-I		& \cite{marguttietal18} \\
iPTF15eqv	& IIb/Ib		& \cite{milisavljevicetal17} \\
PS16op		& SLSN-I		& \cite{marguttietal18} \\
2016ard		& SLSN-I		& \cite{marguttietal18} \\
2016coi		& Ib			& \cite{terreranetal19} \\
2016gkg		& II			& \cite{marguttietal16} \\
2016jae		 & 1a			& \cite{dwarkadas24} \\
2017dwq		& IIn			& \cite{sokolovskyetal17} \\
2017gas		& IIn			& \cite{cc17} \\
2017gir		& IIn			& \cite{canoetal17} \\
2017hcc		& IIn			& \cite{chandraetal22} \\
2018cqj		& Ia			& \cite{dwarkadas24} \\
2018fhw		& Ia			& \cite{dwarkadas23a} \\
2018gk		& IIb			& \cite{boseetal21} \\
2018hti		& SLSN-I		& \cite{andreonietal22} \\
2018ivc		& IIL			& \cite{bostroemetal20} \\
2018zd		& IIn			& \cite{chandraetal18} \\
2019ehk		& Ca-Rich		& \cite{jacobsongalanetal20} \\
2020bvc		& Ic			& \cite{hoetal20} \\
2020eyj.            & Ia-CSM.         & \cite{kooletal23} \\
2021csp.           & Icn		& \cite{perleyetal22} \\
2021gno		& Ca-Rich		& \cite{jacobsongalanetal22} \\
2022ablq		& Ibn			& \cite{pellegrinoetal24} \\
2023ixf		& IIP			& \cite{grefenstetteetal23,chandraetal24} \\
2024ggi		& IIP 		& \cite{marguttietal24a, lutovinovetal24} \\
2024iss		& II			& \cite{marguttietal24b} \\
\hline
\end{longtable}

\noindent{\footnotesize{\textsuperscript{1} SN 1987A data up to day 10433 from \cite{franketal16}. Remaining data from \cite{ravietal24}.}}\\

{As mentioned, all detected X-ray SNe except for one, SN 2012ca, are of the core-collapse variety. SN 2012ca is a Type Ia-CSM SN (\cite{bocheneketal18}. These are a rare and unusual class of SNe (\cite{silvermanetal13} that have an optical spectrum very similar to that of a Type Ia SN, but show the presence of a hydrogen line in the spectrum. The general characteristic of a Type Ia is that it has no hydrogen. The presence of H suggests interaction of the {SNIa} shock wave with a dense, H-rich medium, either formed by mass-loss from a companion star, or by sweeping up the companion material. Upper limits have been listed for other Type Ia-CSM SNe that have been observed but not detected in X-rays, as well as for some Type Ia SNe. The Ia's and Ia-CSMs listed had shown a H-line in their spectrum, indicating strong signs of interaction, and therefore suggesting that they may be good candidates to emit X-rays.}

	When a SN is not detected, the upper limit is generally obtained by computing the maximum number of possible counts in the source region, perhaps by a Bayesian method \cite{kbn91}, and then converting that to a flux using the appropriate column density and emission model. For Type Ia SNe with a H line in the optical spectrum it is difficult to know what column density to use to calculate the upper limit. On the one hand, the presence of a dense CSM (required to produce the H$\alpha$ line, see \cite{hughesetal07,dwarkadas23a,dwarkadas24}) indicates the presence of a large column of material surrounding the SN. Alternatively, the lack of X-ray detection may mean that the SN shock has overrun that column of surrounding material. For the ones listed, in order to maintain some uniformity, we have listed the flux resulting from using a temperature around 5 keV, and the intrinsic (Galactic) column towards the source. The authors have generally listed fluxes using other column densities and temperatures in their respective papers.

Ca-Rich SNe may also arise from the disruptions of low-mass carbon-oxygen (CO) white dwarfs (WDs) by hybrid helium-CO (HeCO) WDs during their merger \cite{zenatietal23}, although this has not been confirmed. 

The {X-ray} luminosities of observed SNe range from a low of 10$^{34}$ erg s$^{-1}$ (not shown) to  a high of 10$^{45}$ erg s$^{-1}$. Below 10$^{36}$ erg s$^{-1}$ there is only one SN, SN 1987A. The early luminosity of SN 1987A  was below 10$^{35}$ erg s$^{-1}$  \cite{schlegel95}. SN 1987A was detected solely because it lies in the Large Magellanic Cloud (LMC), at a distance of approximately 50 kpc. A SN of similar or lower luminosity at extragalactic distances, meaning in a galaxy with a distance much greater than that of the Magellanic clouds, would be too faint to be detected by current X-ray telescopes. This implies that there may exist a large number of low luminosity {X-ray} SNe in galaxies at distances larger than 1 Mpc, with luminosities comparable to {that of} SN 1987A, {but} are too faint to be detected with current instruments.

The initial low {X-ray} luminosity of SN 1987A was due to its evolution in the low density wind of a wind-blown bubble formed by the progenitor star(s) \cite{bl93}. Collision of the SN shock wave with the denser, ionized HII region within the outer shell of the bubble led to an increase in the X-ray emission \cite{cd95}. The emission continued to increase \cite{franketal16} as the SN shock wave collided with the equatorial ring, the equatorial part of the hour-glass shell surrounding the SN \cite{mccray93, mf16}. The shock has now crossed the ring \cite{franssonetal15} and is expanding presumably in a lower density wind outside the ring. The X-ray emission has begun to turn over and decrease \cite{ravietal24}. The reverse shock is moving back into already shocked ring material as well as ejecta, and will begin to contribute significantly to the X-ray emission. The X-ray spectra should show signs of heavier elements in the near future.

Recently, SN 1885A (S Andromedae) in M31 has been claimed to be detected at both radio \cite{1885aradio} and X-ray \cite{1885axray} wavelengths. The X-ray detection was made by combining 45 Chandra HRC-I imaging observations taken between 2001 and 2012. The authors conclude that it has a luminosity of 6$^{+4}_{-3} \times 10^{33}$ erg s$^{-1}$ using an intrinsic column density towards the source of 10$^{21}$ cm$^{-2}$. It is not clear if a 120 yr-old SN should be counted as a SN or a supernova remnant (SNR), and where the dividing line between the two lies. If it is considered a SN, it would be by far the oldest {one} detected, almost 56 years older than the next oldest SN, SN 1941C \cite{sp08,rd20}. It would also be the faintest, being more than an order of magnitude fainter than SN 1987A in its initial stages, although it is about a 100 years older.  Many authors have suggested that SN 1885a is a Type Ia SN. Since these SNe arise from white dwarfs, they are not expected to have a high density surrounding medium, and hence the X-ray flux is expected to be low. However, in a constant density medium, the X-ray flux is expected to increase with time \cite{rd20,ad22}, due to the increase in the volume of the emitting region. This suggests that the density must be extremely low to have such a low flux at an age of 140 years. It is possible that the flux has been underestimated. The column density towards the source was used to calculate the unabsorbed luminosity; if a larger column of material exists around the SN, then this has not been taken into account. It could increase the column density and thereby the unabsorbed luminosity.

At the upper end the brightest SN that has been detected is a superluminous SN (SLSN), SN SCP06F6. The luminosity is on the order of 10$^{45}$ erg s$^{-1}$ \cite{levanetal13}. This source was detected in XMM-Newton observations of 9.9 (MOS1) and 8.22 (MOS2) ks.  The observations were significantly impacted by background flaring, such that data are quoted only in the 0.2-2 erg s$^{-1}$ cm$^{-2}$ band. Strangely, in a subsequent 5 ks Chandra observation 3 months later, no counts were seen, giving an upper limit of 2.5 $\times 10^{44}$ erg s$^{-1}$, lower than the XMM-newton detection. The authors assert this to mean that the source detected by XMM-Newton is transient. However, this would imply a decrease in luminosity of a minimum factor of 2.5, and perhaps as much as a factor of 8, in 3 months. That decrease in luminosity is very high compared to other SNe, and unusual even for SLSNe-I. This is also the only SN in that luminosity {range}. The next brightest confirmed detection is that of SN 2010jl \cite{chandraetal15} with an initial  luminosity at around 40 days of 10$^{42}$ erg s$^{-1}$. Other observations of SNe $> 10^{42}$ erg s$^{-1}$ only {yield} upper limits. This puts SCP06F6 in a category of its own, or makes the luminosity questionable. 

It is interesting to note that the oldest SNe all have luminosities between about 10$^{36}$ and 10$^{38}$ erg s$^{-1}$. It seems that most SNe beyond about 15,000 days tend to cluster in this region on the luminosity-time plot.

\begin{figure}[H]
\begin{center}
 \includegraphics[width=\columnwidth]{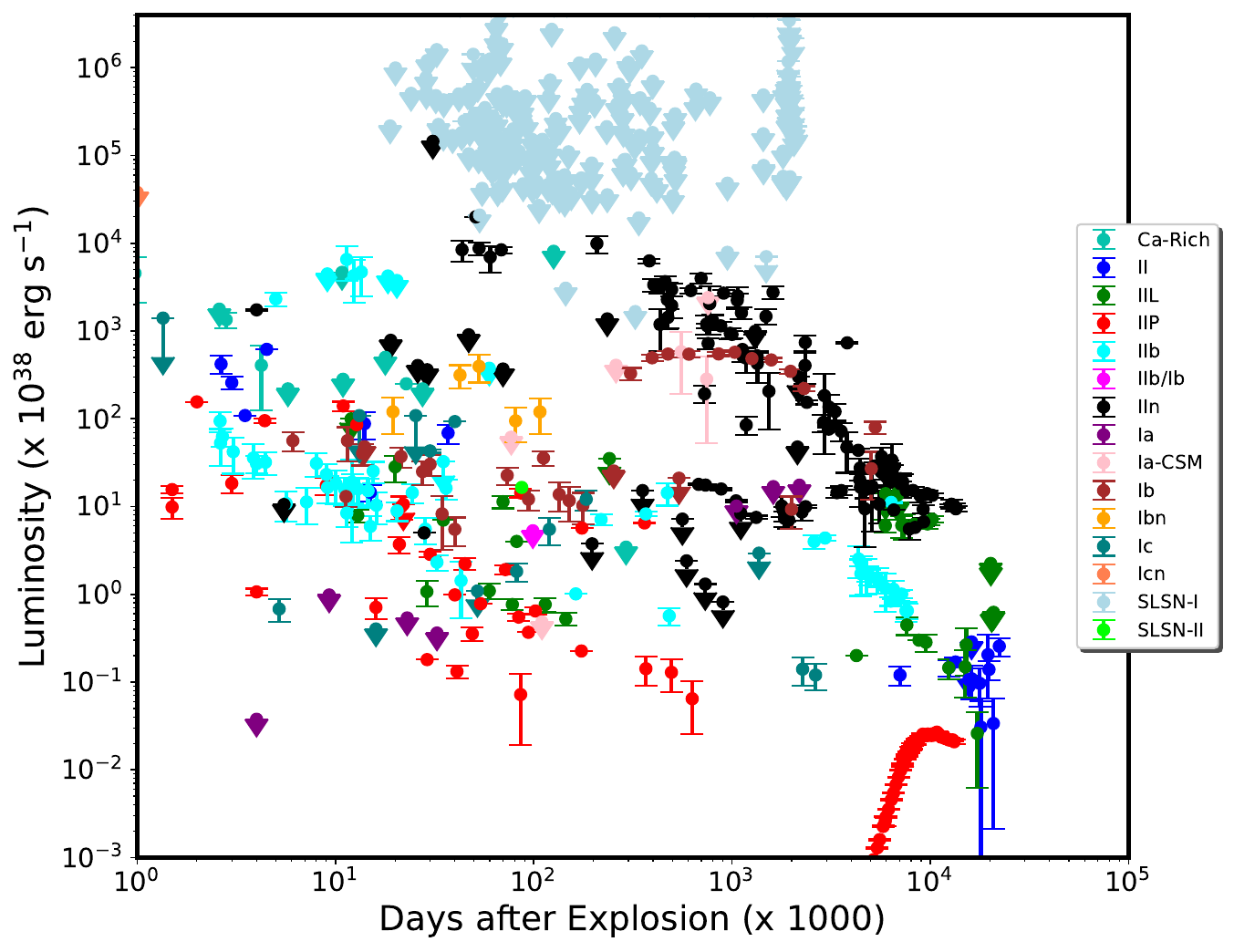} 
 \caption{X-Ray Lightcurves of X-Ray SNe known to the author, grouped by type. SNe that were not detected but have upper limits are plotted with a downward facing arrow.
}
\label{fig:sntype}
\end{center}
\end{figure}

Figure~\ref{fig:sntype} shows all the SNe grouped by type of SN. Only the major types (Type I and Type II) and some major sub-types are listed. There appear to be several more mixed subtypes that appear in the literature, showing characteristics of more than one subtype, such as IIn-P. Unfortunately X-ray observations of such mixed subtypes are not always available. Table~\ref{tab:numtype} provides the number of X-ray SNe by type. In some cases, usually in older SNe, the sub-type is not distinguishable, and they are simply classified as Type II.

\begin{table}[H]
\caption{Number of  SNe of a given type observed in X-rays. \label{tab:numtype}} 
\begin{tabularx}{0.5\textwidth}{LL}
\toprule
\textbf{SN Type}	& \textbf{Number of X-ray SNe} \\
\midrule
IIP & 13 \\
IIL &  7 \\
IIb &  11 \\
Ic &   7 \\
IIb/Ib & 1 \\
Ib &  5 \\
IIn &  19 \\
II   &  10 \\
SLSNe-I & 27 \\
SLSNe-II & 1\\
Ibn & 1 \\
Icn & 1 \\
Ia &  5 \\
Ia-CSM & 5 \\
Ca-Rich & 2 \\
\bottomrule
\end{tabularx}
\end{table}

Almost all the high-luminosity upper limits belong to SLSNe-I. These are superluminous SNe that show no evidence of H in the spectrum. 

{In Figure~\ref{fig:xraysne} it is difficult to distinguish all the SNe. There are 27 SLSNe-I on this list, of which all except 2 have only upper limits. As discussed later, even {for} the two that have been detected in X-rays, the detections or the fluxes may be suspect.}

\begin{figure}[H]
\begin{center}
 \includegraphics[width=\columnwidth]{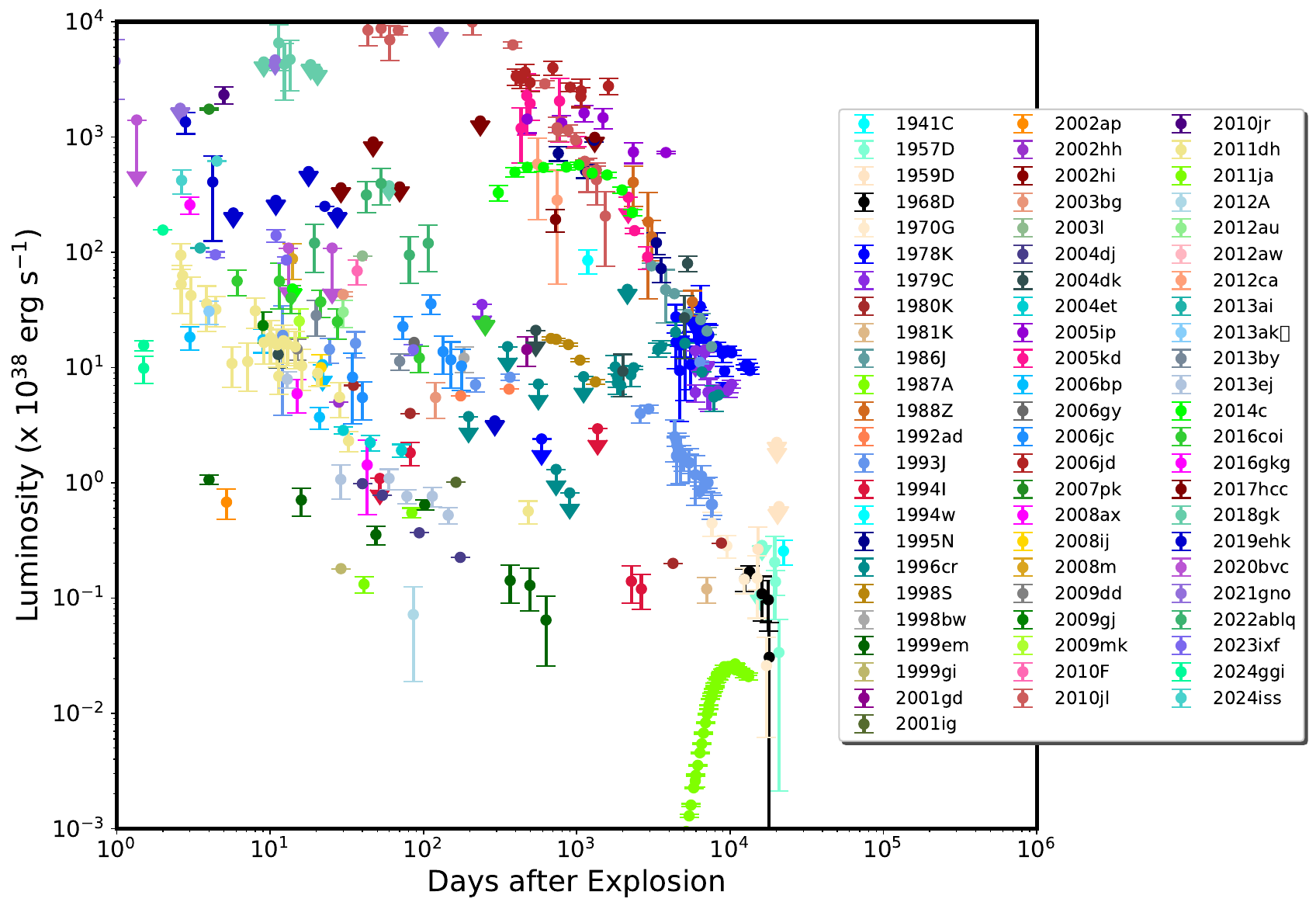} 
 \caption{X-Ray Lightcurves of  X-Ray SNe which have been detected at least one epoch, excluding SLSNe-I. This list includes 70 SNe.
}
\label{fig:sneallshort}
\end{center}
\end{figure}

{In Figure~\ref{fig:sneallshort} we show the lightcurves of all SNe which have at least one X-ray detection. We exclude all the SLSNe-I, as well as other SNe with only upper limits. This leaves 70 SNe with at least 1 detection, making it easier to view the lightcurves of SNe that have been detected.}

\begin{figure}[H]
\begin{center}
 \includegraphics[width=\columnwidth]{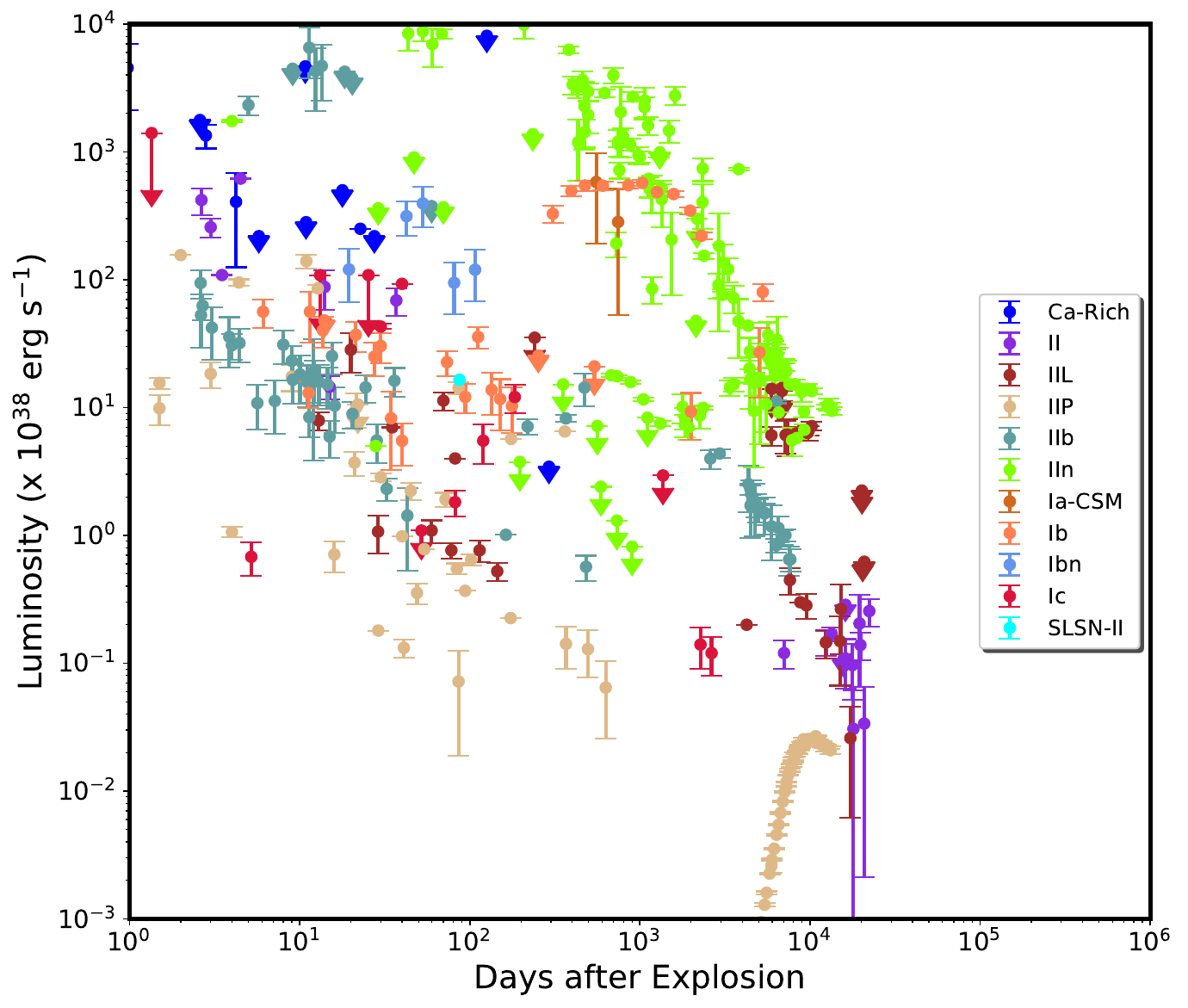} 
 \caption{X-Ray Lightcurves of X-Ray SNe with at least one detction, grouped by type. SNe that were not detected but have upper limits are plotted with a downward facing arrow.
}
\label{fig:sntypeshort}
\end{center}
\end{figure}

{Figure~\ref{fig:sntypeshort} shows the X-ray detected SNe, as in Figure~\ref{fig:sneallshort}, this time grouped by type.  The brightest X-ray SNe generally belong to the Type IIn SN class, shown in green. It is not surprising that this type has the largest number of X-ray detected SNe, 15 in total.  }

{At the lower luminosity end are the Type IIP SNe, in brown, which show the lowest X-ray luminosities. Although IIPs constitute almost half of all core-collapse SNe, in X-rays only 13 have been detected, consistent with their low luminosities. }

{In between the IIPs and the IIns lie the SNe IILs (6 in number), the SNe Ibs (5) and Ic's (6) and the SNe IIb (10). One SN is classified as a Ibn. 10 SNe have unknown subtype, but are known to be Type II. The remaining SNe detected in X-rays include 1 SLSN-II, 1 Type Ia-CSM SNe and 2 Ca-rich SNe.}

\section{X-Ray Emission}
\subsection{X-Ray Spectra of SNe}
\label{sec:spectra}

X-ray emission from young SNe can be of both thermal and non-thermal origin. Non-thermal emission, such as synchrotron or inverse Compton emission, results in a power-law spectrum, modified by the column density, especially at low energies. If the emission is thermal, the spectrum is generally thermal bremsstrahlung modified by the presence of lines, whose flux can be substantial. If lines of various species such as Ne, Mg, Si, S, Ar, Ca, and Fe are present, it is clear that the spectrum is thermal. Most satellites such as Chandra, XMM-Newton and Swift observe in the 0.3-10 keV X-ray band. Above 2 keV there are only a few spectral lines generally present, including those of S, Ca, Ar and Fe, which are not always luminous, and may often not be detected in low counts spectra. In the absence of lines, a high kT thermal spectrum may be mistaken for a non-thermal one. Furthermore, the spectra could be composed of multiple components. The forward shock and reverse shock could each make a contribution to the total X-ray emission. One component may be thermal, the other non-thermal. Thus, although it appears simple in principle to distinguish between thermal and non-thermal spectra, in practice it is much more difficult, due to low counts, and the fact that it is possible for some SNe to have both a thermal and non-thermal emission component. 

In order to get an idea of the emission mechanism from various SN types,  we study the spectra from select SNe of different types. The intent of this paper is not to do an exhaustive study of X-ray spectra, which is beyond the scope of this review. Instead we generally pick one {or two} SNe from a given type, fit the spectra using different models, and determine which provides the best fit. We have used the \textit{vapec} model to represent a thermal model, and the \textit{powerlaw} model to represent non-thermal models. The situation can get more complicated. The emission may be thermal but not in ionization equilibrium, and thus non-equilibrium ionization (NEI) models are needed. However, the ones that are available have their own problems, and are not necessarily suited to model the plasma in young SNe, although they are often used. They all have in-built assumptions such as a planar shock (\textit{pshock}), a Sedov stage model (\textit{sedov}) or a constant temperature with single ionization parameter(\textit{nei} models). The latter is often used to model young SNe. We have used it in our modelling of SN 1979C for the sake of comparison. In all the other cases we use either the \textit{vapec} or \textit{powerlaw} models. 

{We have used the Chandra CIAO software \cite{ciao} version 4.16 and CALDB 4.11 to reduce the Chandra data. Analysis of XMM data was done with xmmsas 20. Fitting of X-ray spectra was carried out with the Sherpa package \cite{sherpa1, sherpa2}.}

We try to pick SNe which have at least a few hundred counts in their spectra, which is not always easy, given that many observed SNe have few counts. Nevertheless, our study provides an idea of the spectrum and what the best fit looks like. In some cases different models fit equally well, and it is difficult to distinguish the best fitting model based on the spectra alone. 

\begin{itemize}
\item {\em Type IIn SNe:} Type IIn spectra are {generally easy to fit}. Since they are the brightest type of SNe {detected in X-rays}, the spectra usually have sufficient counts to provide a good fit.  IIns that have been observed in X-rays thus far \cite{houck05,baueretal08,chandraetal09a,chandraetal12,chandraetal15,drrb16,thomasetal22} display thermal spectra, with clearly identifiable lines, generally of $\alpha$ elements. The spectra of SN 1986J and SN 1995n shown in Figure~\ref{fig:iinspectra} are well fit with \textit{vapec} models , with distinct line emission visible. Sometimes more than one thermal component may be needed. Type IIn SNe have some of the highest X-ray luminosities amongst all SNe (see Figure~\ref{fig:sntypeshort}). Thermal X-ray emission is directly proportional to the square of the density. The high luminosity therefore implies that they are expanding in media of very high densities. Given the high density of the material surrounding it, many Type IIn also show high absorption, evidenced by a high column density. SN 2010jl had an initial column density of 10$^{24}$ cm$^{-2}$ \cite{chandraetal15}.

\cite{ci12} have postulated that SN shocks expanding in very dense environments may show inverse Compton X-ray emission. \cite{chandraetal15} do find that a power-law component is needed to better fit the 2011 October and 2012 June observations of SN 2010jl, which they attribute to a cooling shock. The presence of inverse Compton emission has not been convincingly demonstrated in Type IIns. Conversely, it must be mentioned that Type IIns have not been detected in the first month. The earliest detection is that of SN 2010jl at around 40 days after explosion\\

\begin{figure}[H]
\begin{center}
 \includegraphics[width=\columnwidth]{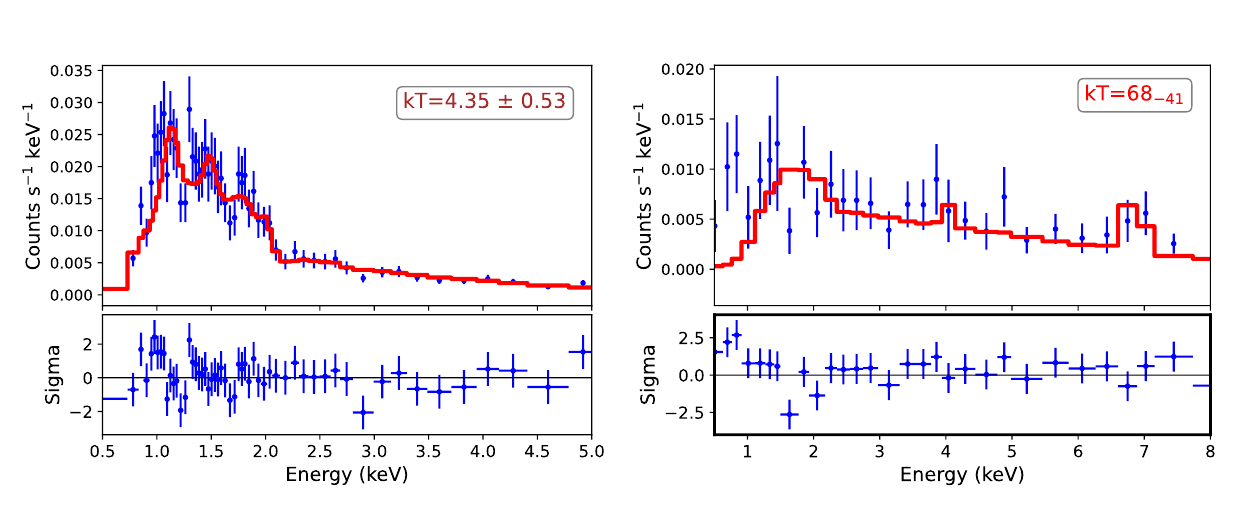} 
 \caption{[Left] Chandra spectrum (blue) and fit (red) of SN 1986J, ObsID 794. Lines of Mg and Si are visible in the fit. [Right] The XMM PN spectrum (black) and fit (red) of SN 2005kd, XMM ID 0410581101. In this case the Fe K line at around 6.7 keV, and a Ca line at $\approx$4 keV, are both apparent. The fit is consistent with that in \cite{drrb16}. In both of these Type IIns, the spectra are clearly thermal.
}
\label{fig:iinspectra}
\end{center}
\end{figure}

\item {\em Type Ib/c SNe:} Type Ib/c SNe have no H (Ib) and no H or He (Ic) in their optical spectra. Their progenitors are thought to be stars which have had the H and sometimes He envelopes stripped off. Ib/c SNe are therefore sometimes grouped together as `stripped envelope SNe', along with the more luminous SLSNe-1. The progenitors are suspected to be Wolf-Rayet (W-R) stars which have lost their H, and sometimes He, envelopes. The progenitor star may be in a binary system where mass from the progenitor star is stripped off by a companion.  

\cite{cf06} have suggested that the X-rays from Type Ib/c SNe arise via a non-thermal mechanism, either inverse Compton or synchrotron. The emission is likely inverse Compton in the first month. This will be followed by synchrotron emission at later times, as long as the shock is cosmic-ray dominated, and the electron energy spectrum flattens at high energy.

Herein we have chosen to study the X-ray emission from SN 2003L, one of the SNe mentioned by \cite{cf06}.  In Figure~\ref{fig:03L_fig} we show two fits to the spectra of SN 2003L (Chandra ObsID 4417). The left panel shows a fit with a thermal \textit{vapec} model and an absorption component described by the \textit{tbabs} model; in the right panel the data are fitted with a \textit{powerlaw} model. Due to the low count statistics, the reduced ${\chi}^2$ for both are much smaller than one, although the \textit{vapec} fit returns a somewhat higher reduced ${\chi}^2$ compared to the \textit{powerlaw} model. The thermal fit is comparable to the fit carried out by \cite{soderbergetal05}. The \textit{powerlaw} fit is different from that in \cite{soderbergetal05}, and returns a higher flux, because we did not restrict the column density to be the Galactic column density as in \cite{soderbergetal05}. We find that the higher column density provides a better fit. The value of the photon index $\Gamma$ is high, but so is the uncertainty. At the lower end, this value of $\Gamma$ may be consistent with synchrotron or Inverse Compton emission with an electron spectral index close to 3.  The fit alone does not enable to distinguish between these possibilities. As shown later, assuming that the X-rays are arising due to thermal bremsstrahlung may require exceptionally high densities, and therefore mass-loss rates, which are not normally associated with W-R star winds.\\

\begin{figure}[h]
\begin{center}
 \includegraphics[width=\columnwidth]{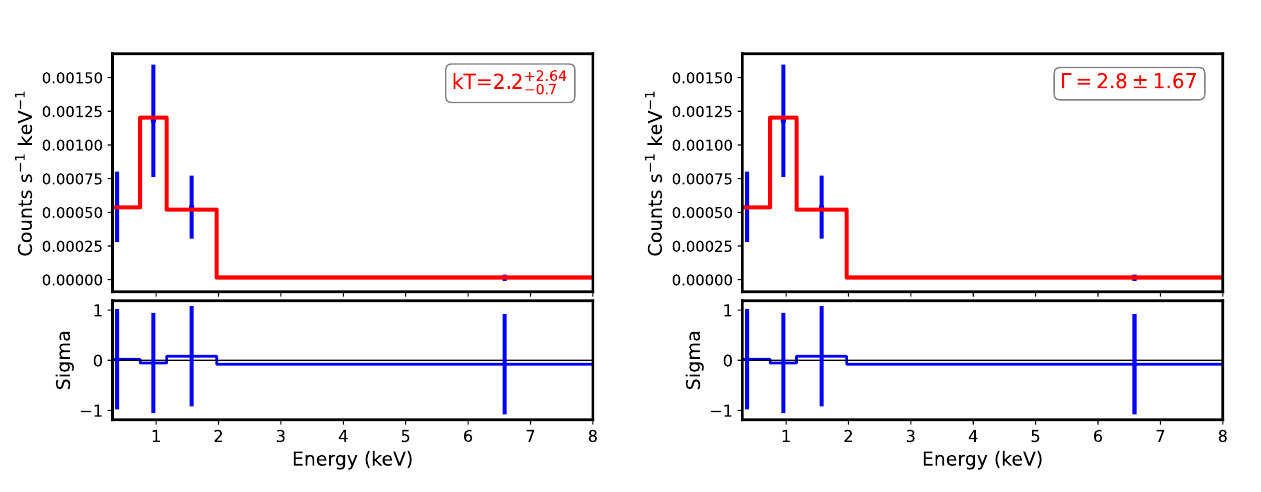} 
 \caption{[Left] A thermal fit to the Chandra spectrum of SN 2003L, a Type Ibc SN. [Right] A powerlaw model fit to the same data.
}
\label{fig:03L_fig}
\end{center}
\end{figure}

\item {\em Type IIb SNe:} Type IIb SNe have a very small amount of Hydrogen in their optical spectrum, somewhere in between normal Type II,  which have a large amount of H, and the Type Ib, which show no H. The prototype Type IIb SN, SN 1993J, shows a thermal X-ray spectrum, with distinct lines of various elements clearly present (\cite{chandraetal09b,ncf09,dbbb14}. In Figure~\ref{fig:93j} we show the 2000 Chandra spectrum of SN 1993J (ObsID 735). The spectrum is clearly thermal, and is fitted well by two \textit{vapec} models. Lines of N, O, Mg and Si are detected. \cite{swartzetal03} have fitted it with three thermal models, but that complexity was not needed herein, as we simply wish to demonstrate the thermal nature of the spectrum.

\begin{figure}[h]
\begin{center}
\includegraphics[width=0.75\textwidth]{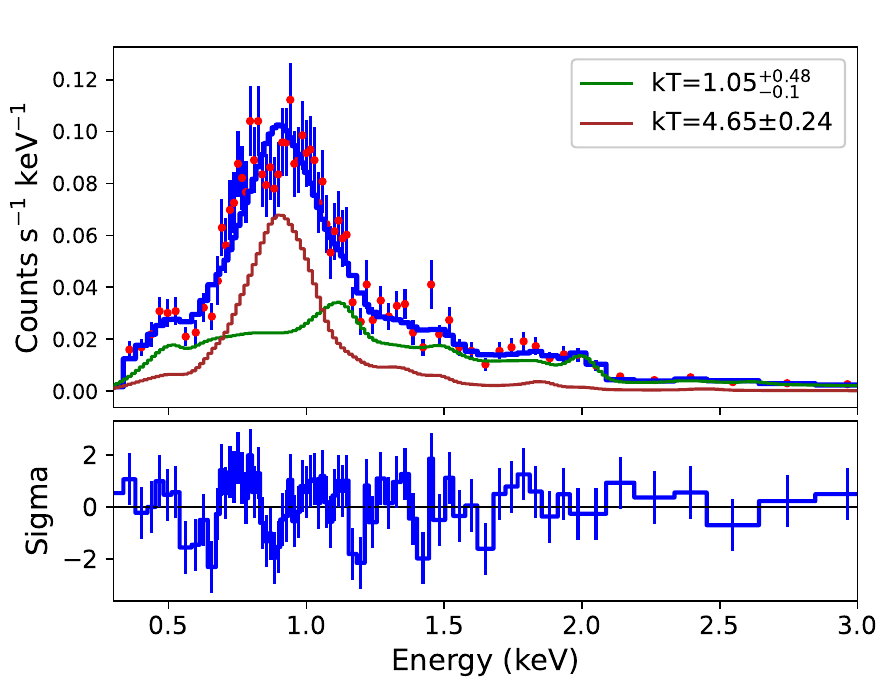}
\end{center}
\caption{ The Chandra spectrum of SN 1993J in the year 2000 (ObsID 735).  }
\label{fig:93j}
\end{figure}

\cite{cs10} have suggested that there is another class of Type IIb SNe with compact progenitors (cIIb), whose X-ray emission may be non-thermal. They cite SNe 1996cb, 2001ig, 2003bg, 2008ax, and 2008bo in this category. SN 2001ig has been observed with Chandra, but the 25ks observation  (ObsID 3495) yields  low counts (32, with 17 in the background), making it difficult to fit the spectrum with any model and categorize the emission. A subsequent spectrum (ObsID 3496) returns even lower counts. Even when the two spectra are combined (not necessarily a good idea as the flux may be decreasing in time), the counts are too low to yield a good fit. 

SN 2003bg was classified by \cite{soderbergetal06} as being a Type Ic SN that transitioned to a Type II. However, \cite{mazzalietal09} asserted that it was a broad-lined Type IIb SN with Hydrogen. \cite{cs10} have considered it a Type IIb. The SN was observed in March 2003 for 50 ks with Chandra. The observation yielded a large number of counts ($>$ 600), making it possible to fit the spectrum reasonably well. Figure~\ref{fig:03bg} shows fits to the data using both a thermal and a power-law model.  Both models fit equally well, with a similar reduced ${\chi}^2$, making it difficult to distinguish between the two. The \textit{vapec} model fit is subtly improved if an excess at 3.9 keV is assumed to be the Ca XIX line. Thawing Ca gives a good fit to the excess, and suggests that the emission mechanism is thermal.

\begin{figure}[h]
\begin{center}
\includegraphics[width=1.\textwidth]{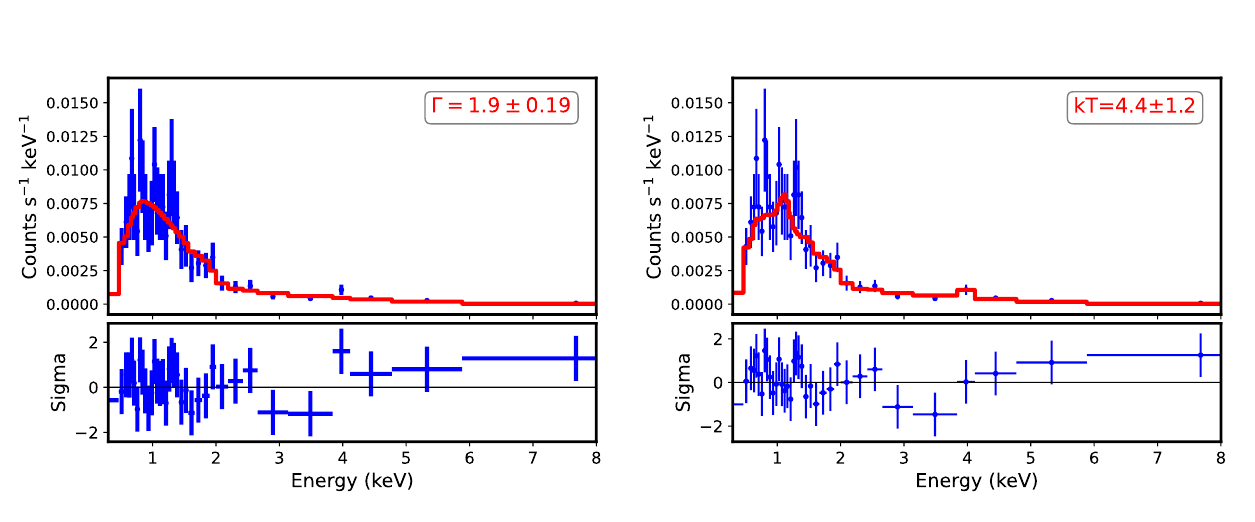}
\end{center}
\caption{ Chandra spectrum of SN 2003bg (ObsID 3870). The data are in blue, fit is in red. [Left] A powerlaw fit, with photon index 1.9 $\pm$ 0.19. [Right]  A \textit{vapec} fit, with temperature 4.4 $\pm $ 1.2 keV, and Ca thawed to fit the excess at $\approx$ 4 keV, assumed to be a  Ca line.}
\label{fig:03bg}
\end{figure}

Tentatively, we can identify the \textit{vapec} fit to be better. This is consistent with \cite{pl03b} who fitted both power-law and thermal models, but prefer the thermal fit. Our fit parameters are also consistent with theirs, although they did not consider a Ca line.

As one example of a cIIb SN, \cite{cs10} cite the example of SN 2008ax. They claim that \cite{romingetal09} found that the SN decreased in X-ray luminosity by a factor of about 4 from the first month to the second month, which is consistent with the decline expected for a SN whose emission goes from an inverse Compton component near optical maximum to a synchrotron component a month later. There are problems with this assumption. Firstly, \cite{romingetal09} were forced to extract the counts from a region < 5$^{\prime\prime}$, whereas the Swift telescope has a half-power diameter of 18$^{\prime\prime}$. Thus it is doubtful that the counts are accurate. More importantly, \cite{romingetal09} then calculated the fluxes using a thermal plasma spectrum with kT=10 keV. These  fluxes were subsequently used by \cite{cs10} to justify the existence of a non-thermal component, which is plainly inconsistent. 

\cite{smk19} computed binary models for the progenitors of Type IIb SNe using MESA. They did not find two classes of Type IIb progenitors. From our exploration of 2003bg, we cannot say that two classes  of IIb SNe exist, with a difference in their X-ray emission. It is possible that most, if not all, IIb's are fit with a thermal model. Deeper observations of many Type IIb SNe may be able to further probe this statement.\\

\item {\em Type IIL SNe:} Type IIL SNe are characterized by a linear decline in their optical lightcurve. There are not many observations of Type IIL SNe in X-rays with high counts that would enable us to characterize the X-ray emission mechanism. SN 1979c was a Type IIL that was observed in X-rays. Its spectrum was better fitted with two thermal  components according to \cite{immleretal05}. While their initial fit indicated a  column density higher than the Galactic column, they assert that using the Galactic column density towards the source gives a similar result. \cite{plj11} do manage to fit the spectrum with two thermal components and a Galactic column density. However, they also find that the spectrum can be equally well fit with a thermal component and a non-thermal \textit{powerlaw} component.

 We have therefore attempted several different fits to the Chandra spectrum of SN 1979C, obsID 6727. In Figure~\ref{fig:79c_fig} we show various fits, including a single-temperature \textit{vapec} model (top left), a non-equilibrium ionization model \textit{vgnei} (top right), two \textit{vapec} components (bottom left), and a \textit{vapec}  component plus a \textit{powerlaw} component similar to \cite{plj11} (bottom right).  The application of an f-test reveals that the \textit{vapec+powerlaw} component is marginally preferred (at the 95.5 \% level) to the single thermal component. This fit is consistent with that in \cite{plj11}.

This example illustrates the difficulty in identifying the spectral components, and thereby emission mechanism, from a SN. It is possible that SN IIL may show multiple components. On the other hand, \cite{plj11} have suggested that SN 1979C may harbor a central black hole, which could explain its almost constant {X-ray} flux. This could be the source of the non-thermal component. This would make SN 1979c a special case, and not representative of the entire group of IILs. We have not calculated the flux from all the available spectra of SN 1979c, and cannot comment on the assertion that the SN harbors a black hole. However, it appears most likely that SNe IIL may require at least one thermal component.\\

\begin{figure}[H]
\begin{center}
 \includegraphics[width=\columnwidth]{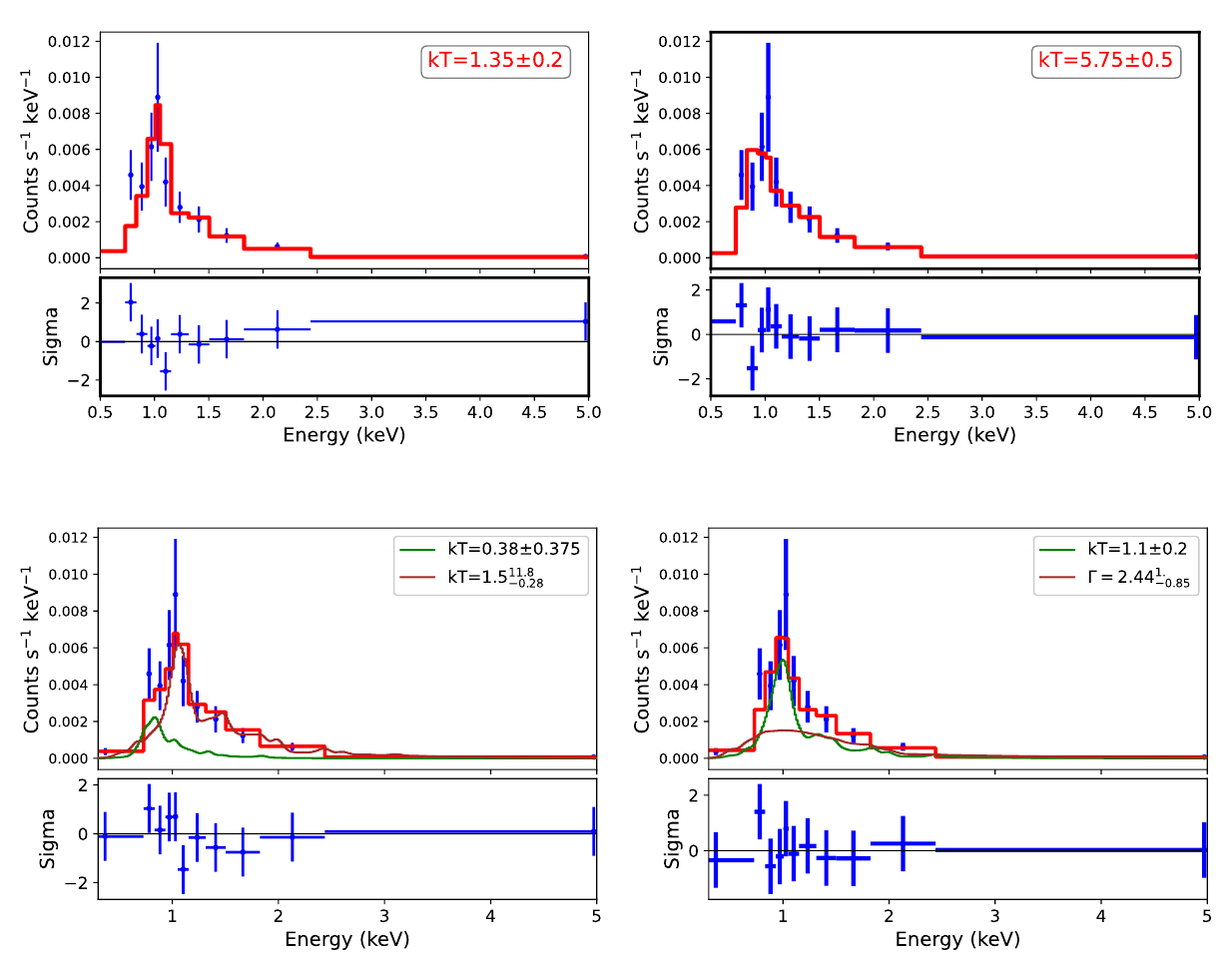} 
 \caption{[Top Left] A thermal \textit{vapec} fit to the Chandra spectrum of SN 1979C (ObsID 6727), a Type IIL SN. [Top Right] A non-equilibrium ionization \textit{vgnei}  model fit to the same data. [Bottom Left] Two \textit{vapec} component fit. the two components are also shown. [Bottom Right] A \textit{vapec+powerlaw} model component fit. For the two-component models, the individual components are also shown. In all cases, data are in blue, fit in red.
}
\label{fig:79c_fig}
\end{center}
\end{figure}

\item {\em Type IIP SNe:} Type IIP SNe show a plateau in their optical lightcurve. They are generally the least luminous of all X-ray SNe. Therefore it is difficult to find examples that have sufficient counts to be able to fit the spectrum accurately and determine the emission model. Initially, the spectra of IIPs were fit with whatever model was deemed appropriate. \cite{pooleyetal02} fit the spectrum of the Type IIP SN 1999em with both a thermal bremsstrahlung model and a MEKAL model, i.e. thermal emission models.

In 2006, \cite{cfn06} studied the X-ray and radio emission from a few Type IIP SNe. They found that both thermal and non-thermal models could potentially match the X-ray emission.  Thermal models would dominate when the density (i.e. ratio ${\mdot}_w/v_w$) is high, whereas inverse Compton may dominate if the ratio was low. If reverse shock cooling became important, especially at early epochs, then the inverse Compton component could also dominate.

In order to investigate the X-ray spectrum, we have employed the data from SN 2004et, Chandra obsID 4631. In Figure~\ref{fig:04et_art} we show fits to the data using both a \textit{powerlaw}  as well as a \textit{vapec} model. Both fits appear to fit equally well, and give similar values for the reduced ${\chi}^2$ statistic. The power-law fit is consistent with synchrotron emission with an electron spectral index $\approx 3$, which has been suggested by \cite{cfn06}. However, there is no reason to choose this over the vapec fit. The temperature in the \textit{vapec} fit is outside the range of Chandra, and therefore unconstrained. Our values for the column density and temperature, or power-law index,  are consistent with the results of \cite{misraetal07}; therefore it is not surprising that the fluxes are also consistent.  Similar to \cite{misraetal07}, our results for this observation are inconsistent with the work of \cite{rhoetal07}. We are unable to explain the {source of} this discrepancy.

Overall it is difficult to choose between the two models based purely on the X-ray spectral fits. This SN is one of the few that has a reasonable number of counts ($\approx$ 250) in the X-ray regime; most SNe IIP have even fewer counts. This makes it difficult to determine the emission mechanism for Type IIP SNe accurately.

\begin{figure}[H]
\begin{center}
 \includegraphics[width=\columnwidth]{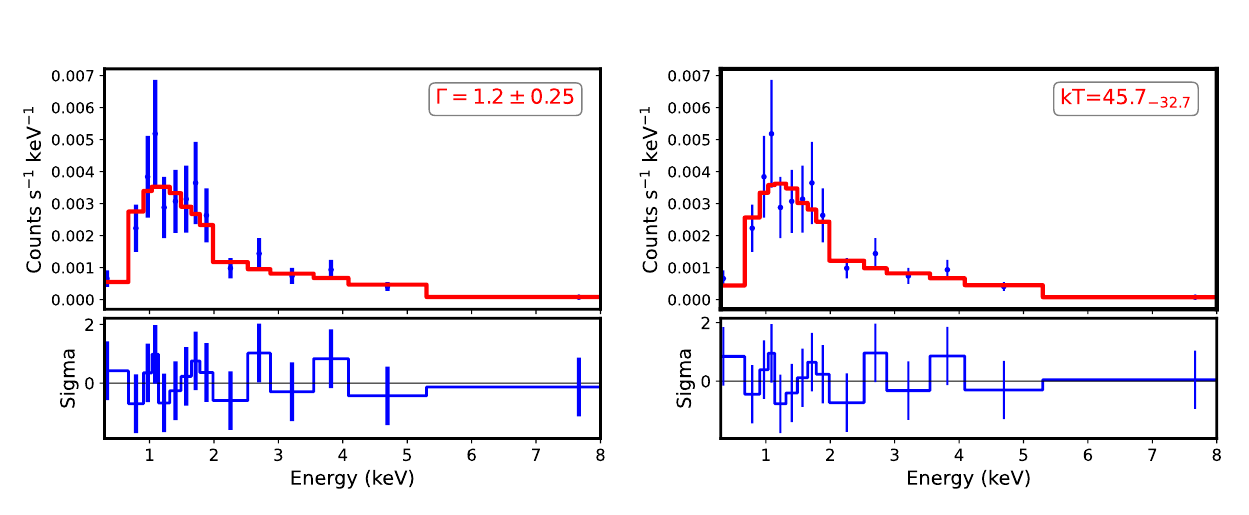} 
 \caption{[Left] \textit{Powerlaw} fit to the spectrum of SN 2004et, ObsID 4631. [Right] \textit{vapec} fit to the same spectrum. In all cases, data are in blue, fit in red.
}
\label{fig:04et_art}
\end{center}
\end{figure}

\end{itemize}

\section{X-ray luminosity and Mass-loss rates:} 
\label{sec:mdot}

The time evolution of the X-ray luminosity of a SN can be related to the density structure of the surrounding medium \cite{flc96, dg12}, which in the case of core-collapse SNe is formed by mass-loss from the massive progenitor star. We use the \cite{chevalier82a,cf94} description for a SN shock wave evolving in a self-similar spherically symmetric manner. The SN ejecta density  ${\rho}_{SN}$ is described as  ${\rho}_{SN} = A\,v^{-n} \, t^{-3}$, where $A$ is a constant. The circumstellar medium (CSM) into which the SN evolves has a density profile ${\rho}_{CSM} = B\, r^{-s}$, where B = $\mdot / (4 \pi v_w) $ for a wind with constant mass-loss parameters ($s=2$), and $B$=constant for a constant density medium ($s=0$).   The interaction of the SN ejecta with the surrounding medium can then be described by a self-similar solution for the expansion of the contact discontinuity R$_{CD}$, with R$_{CD} \propto t^m$, m=(n-3)/(n-s).

Core-collapse SNe are expected to evolve in a wind medium formed by mass-loss from the progenitor. A value $s=2$ indicates a wind with constant mass-loss rate and velocity. If the parameters are not constant, then $s$ deviates from 2. Type Ia SNe are expected to arise from white-dwarf stars, which are not expected to lose mass (although mass-loss from a companion main-sequence star is possible). Thus they are generally expected to evolve in a constant density medium, $s=0$.

It is important to note that a freely expanding wind cannot directly interact with the surrounding medium, since such a configuration would be hydrodynamically unstable. Instead, these winds form wind-blown nebulae surrounding the massive stars. The structure of such a wind-blown region was first elucidated by \cite{weaveretal77}. Going outwards in radius, we find (1) A freely flowing wind. (2) A shocked wind region. (3) The shocked ambient  medium. (4) The unshocked surrounding medium. Regions (1) and (2) are separated by a wind-termination shock, regions (2) and (3) by a contact discontinuity, and regions (3) and (4) by an outer shock, which can be shown to almost always be radiative. Ionizing photons from a hot star can create an ionized HII region interior to the contact discontinuity, as shown by \cite{ta11,dr13,dwarkadas22,dwarkadas23b}. A large fraction of the volume of these `bubbles' consists of the low density, high pressure region of shocked wind. 

Close in to the star, we find a freely flowing wind. Thus it is generally assumed that core-collapse SNe initially expand within a freely flowing wind. Some young SNe, however, do show evidence of expanding within a wind-blown bubble. The classic example is SN 1987A.  Increasing X-ray emission from the SN was initially attributed to the interaction with the ionized HII region \cite{cd95}, and subsequently with the equatorial ring, or the waist of an hour-glass shaped wind bubble surrounding the SN.  In the last two decades, other SNe have also shown evidence of increasing X-ray emission, and of possible evolution in a bubble, including SN 2004dk \cite{pooleyetal19}.

{X-Ray Luminosity:} The thermal X-ray luminosity can be written as $L_x \propto n_e^2 \Lambda V$, where $n_e =$ number density of ambient wind, $\Lambda$ is the cooling function and $V$ the volume of emitting material.

Using the above description, the X-ray luminosity can be expressed in terms of the parameters $n$ and $s$ as \cite{flc96, dg12}:

\be 
\label{eq:lumgen}
L_x \sim \; t^{-(12-7s+2ns-3n)/(n-s)}
\ee

\begin{figure}
\begin{center}
 \includegraphics[width=\columnwidth]{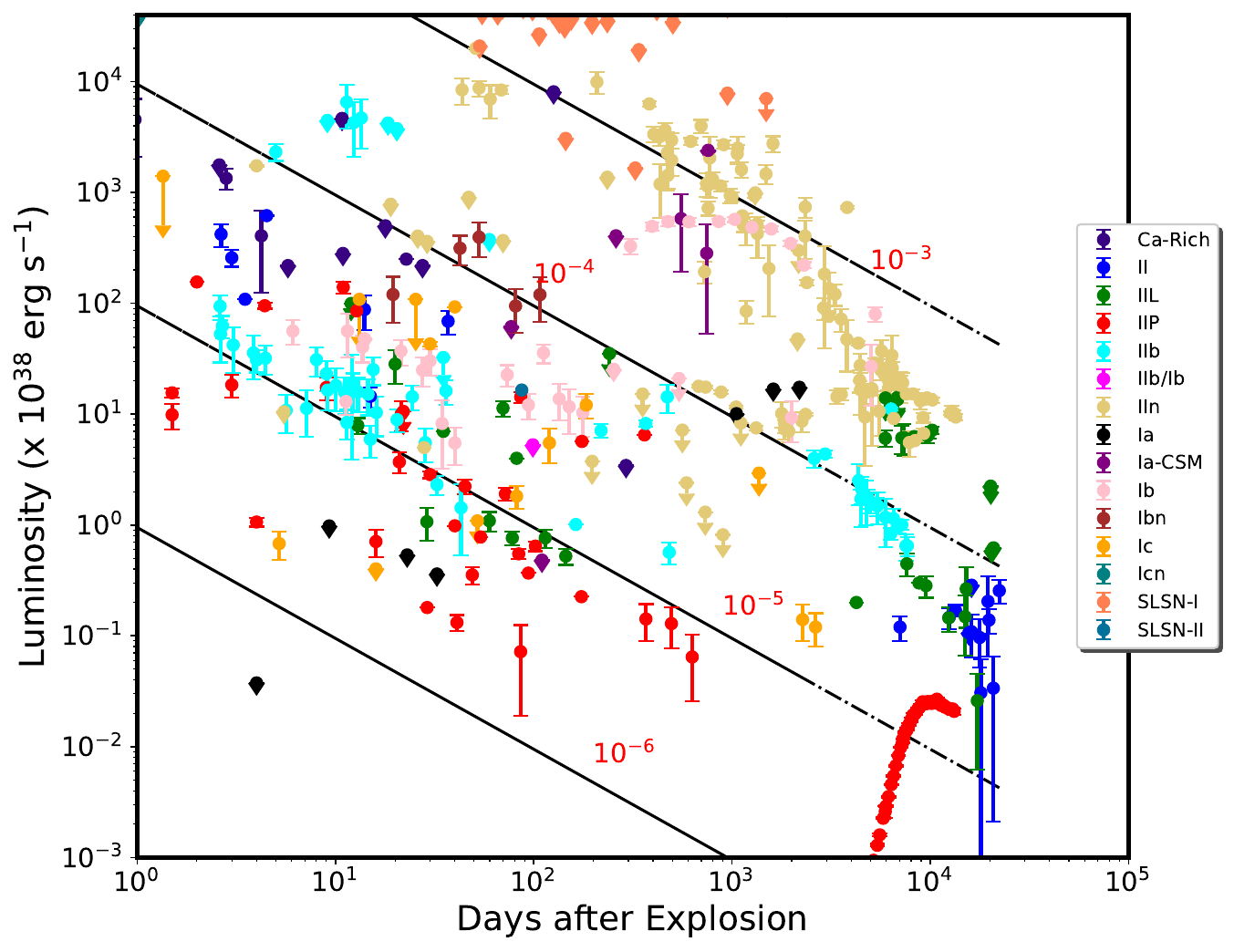} 
 \caption{X-Ray Lightcurves grouped by type. The {dot-dash} lines represent lines of constant mass-loss, with  t$^{-1}$ slope. Mass-loss rates are given in $\msun$ yr$^{-1}$, assuming a wind with constant parameters (density $\propto$ r$^{-2}$), v$_{\rm wind}$ = 10 km s$^{-1}$. The electron temperature is assumed to be 10\% of the ion temperature behind the shock \cite{cf03}.
}
\label{fig:lcmdot}
\end{center}
\end{figure}

For $s$=2, the emission decreases with time as $L_x \propto t^{-1}$. This assumption has been widely used to describe X-ray light curves of SNe, and compute the mass-loss rate of the wind $\dot{M}$ \cite{pooleyetal02, immleretal02, il03, ik05, immler07, ki07, sp08, milleretal10}. However, as shown in Figure~\ref{fig:lcmdot}, the X-ray lightcurves of many SNe do not follow this trend, suggesting that the   the density structure of the CSM does not necessarily correspond to a steady wind.

X-ray observations rarely cover the entire X-ray range. Modern X-ray satellites such as Chandra, XMM-Newton and Swift have an approximate  range 0.3-10 keV. While NuStar extends to 79 keV, it has observed very few SNe. The luminosity in a given energy band with E $<< kT_{sh}$ goes as \cite{flc96}:

\be 
\label{eq:lumband}
L_x(E) \sim \; t^{-(6-5s+2ns-3n)/(n-s)} \sim \; t^{-\beta}
\ee

\noi 
Compared to eqn.~\ref{eq:lumgen}, note the expected flatter time dependence in a given X-ray band. For $s=2$, $L_x \sim t^{-(n-4)/(n-2)}$. The index $\beta$ is always $< 1$. For $n > 5$, as required \cite{chevalier82a}, $0.33 < \beta < 1$.

A radiative reverse shock has luminosity $L_x \sim \; t^{-(15-6s+ns-2n)/(n-s)}$ \cite{flc96}.

Expressions for other cases are given in \cite{flc96, dg12}.

{In the framework of the self-similar solution}, the X-ray luminosity for a SN shock wave expanding into a constant-parameter wind ($s=2$) can be written in terms of the wind parameters as (\cite{cf03}):

\begin{equation}
\label{eq:xraylum}
L_i = 3 \times 10^{39} g_{ff} C_n \left(\frac{{\mdot}_{-5}}{{v_w}_{10}} \right)^2 \left(\frac{t}{10 \,{\rm d}} \right)^{-1} {\rm ergs\; \;s^{-1}}
\end{equation}

\noindent
where  $g_{ff}$ is the Gaunt factor, $C_n = 1$ for the circumstellar shock and $(n-3)(n-4)^2/4(n-2)$ for the reverse shock. Note that since the reverse shock is expanding into a higher density medium as compared to the forward shock, its velocity is generally lower, and therefore the temperature of the emission is lower. The higher density can lead to a higher flux from the reverse shock. ${\mdot}_{-5}$ is the mass loss rate of the wind into which the SN is expanding, in units of 10$^{-5}$ $\msun$ yr$^{-1}$, ${v_w}_{10}$ is the wind velocity in units of 10 km s$^{-1}$ (characteristic of red supergiants), and $t$ is the time in days. The mass loss rate and wind velocity are assumed constant, leading to a density in the wind medium that decreases as $r^{-2}$, and an X-ray luminosity that consequently decreases as $t^{-1}$ \cite{flc96,dg12}.  {The CSM is assumed to be spherically symmetric}. Electron-ion equilibration behind the shock front is assumed, probably not true for young SNe, except when the density is very high. However, the equation can be easily modified to account for any specified ratio of electron to ion temperature (T$_e$/T$_i$), simply by multiplying the right hand side by (T$_e$/T$_i)^{1/2}$.

If we exclude SN SCP06F6 as well as the early observations of SN 1987A, and the upper limits to almost all the SLSNe-1, then the remaining observed population of X-ray SNe, which encompasses all but 2 (3 if we count SN 1885A) of the observed SNe, lies mainly between about 10$^{36}$ and 10$^{42}$ erg s$^{-1}$. Thus in general the majority of observed SNe span about 6 orders of magnitude in luminosity.
Figure~\ref{fig:lcmdot} modifies Figure~\ref{fig:sntype} by excluding superluminous SNe and the early X-ray observations of SN 1987A. 

Inverting equation~\ref{eq:xraylum}, we overplot lines of X-ray luminosity at a  constant mass loss rate, with T$_e$/T$_i$ = 1/10. These lines of constant mass-loss rate are for emission from the forward shock. In the case of the reverse shock, for a representative value of n=9 and s=2, the luminosity would be larger by a factor of $\approx 5.36$.

Figure~\ref{fig:lcmdot} can be used to estimate the mass-loss rate of an individual SN. The accuracy depends on several factors, specifically on how closely the light curve follows the $t^{-1}$ luminosity decline. If the wind is steady and the X-ray luminosity is decreasing sufficiently close to $t^{-1}$, these rates are similar to more accurately calculated ones. However, in cases where the lightcurve departs strongly from a $t^{-1}$ slope, as for IIns at late times, the rates can deviate considerably from the values shown here. If the luminosity does not decrease as $t^{-1}$, {the most likely explanation is that} the wind properties are not constant, and consequently that the mass-loss rate is a function of time, or radius. 

It should be noted that the defining parameter here is $\mdot/v_w$. These mass-loss rates are for a particular value of the wind velocity, 10 km s$^{-1}$. If the wind velocity is higher, the mass-loss rate must be proportionally increased.

Although the rates are approximate, the plot reveals several interesting characteristic of the mass-loss rates of various SN types, and thereby of their progenitor stars.

\begin{itemize}
\item {\em Type IIP SNe:} Type IIP SNe are known to have red supergiant (RSG) progenitors. As seen from Figure~\ref{fig:lcmdot}, these SNe tend to have the lowest mass-loss rates (in red), lower than about 10$^{-5}$ $\msun$ yr$^{-1}$ in most cases. It is known that red supergiant can have higher mass-loss rates, up to 10$^{-4}$ $\msun$ yr$^{-1}$ \cite{mj11}. The fact that the resulting SNe generally had mass-loss rates lower than about 10$^{-5}$ $\msun$ yr$^{-1}$ led \cite{dwarkadas14}  to conclude that the progenitors of Type IIP SNe were red supergiants with initial mass $< 19 \,\msun$. This result is in agreement with results derived via direct optical measurements (\cite{smartt09, smartt15}, as well as theoretical calculations (\cite{ekstrometal12, sukhboldetal16}. They all seem to indicate that RSGs above about 19 $\msun$ do not explode to form Type IIP SNe. It is true that this value is approximate; {however, the important point here is} that as one moves up in the plot, the X-ray luminosity increases, and therefore the density of the surrounding medium must increase if the emission is thermal. If the surrounding medium is formed by mass-loss from the progenitor star, then the mass-loss rate must increase as we move up the plot. And since in general the mass-loss rate of massive stars is proportional to the mass, the mass of the star must increase as we move up the plot. Thus this clearly indicates that Type IIPs arise from the lower mass end of the massive star population.

In the last couple of years, two Type IIP SNe, SN 2023ixf and SN 2024ggi, were both detected at relatively close distances (< 10 Mpc), and consequently observed in X-rays just days after their discovery. Optical observations suggested a mass-loss rate $\approx$ 10$^{-2}\, \msun$ yr$^{-1}$ for SN 2023ixf. Yet the X-ray luminosity did not exceed a few times 10$^{40}$ erg s$^{-1}$, which is clearly low for the suggested mass-loss rate. The reason for this discrepancy is not clear, although an asymmetric medium, or a radiative shock, has been suggested. It is, however, in accordance with the low X-ray luminosities of Type IIP SNe. 

We note that this value of the maximum mass-loss rate and hence the mass assumes that the emission is thermal. If the emission is non-thermal, that must mean that the mass-loss rate of IIPs must be much lower, such that the non-thermal emission dominates. This would make the maximum mass-loss rate, and thus maximum progenitor mass, even lower. In general whether the emission is thermal or not will depend on the density of the surrounding medium, and thereby on the mass-loss rate, if the surrounding medium is formed via mass-loss.  The X-ray luminosity of Inverse Compton emission, for the case where the energy spectral index $p=3$, takes the form \cite{cfn06}:

\noindent
\be
\frac{ dL_{ IC}}{ dE} \approx 8.8 \times 10^{39}\, {\epsilon}_{ r} \, { \gamma}_{min} \,E_{keV}^{-1}\, \left[\frac{\dot{M}_{-5}}{ v_w}_{10}\right]\,V_{s4} \left[\frac{{\rm L}_{ bol}( t)}{10^{42}\; { ergs}\; { s}^{-1}}\right] { t}_{10}^{-1}\; { erg}\; { s}^{-1}\;{ keV}^{-1}
\label{eq:ic}
\ee

\noindent
where ${\rm L}_{\rm IC}$ is the Inverse Compton Luminosity, $\gamma_{\rm min}$ is the minimum Lorentz factor of the relativistic electrons, $ V_{s4}$ is the forward shock velocity in units of 10$^4$ km s$^{-1}$, ${L}_{ bol}$ is the bolometric luminosity of the SN, and ${\epsilon}_{ r}$ is the ratio of relativistic energy density to thermal energy density. This expression can be modified for an arbitrary value of $p$, and for evolution in arbitrary media~\cite{marguttietal12}.

Thermal emission depends on the square of the density, i.e. on ${[\dot{M}/v_w]}^2$ (equation~\ref{eq:xraylum}), whereas the inverse-Compton emission depends on a single power of  ${[\dot{M}/v_w]}$ (equation~\ref{eq:ic}). Both expressions have a similar time dependence. Thus it is clear that for high mass-loss rates (with a fixed wind velocity of 10 km s$^{-1}$, characteristic of RSG stars), we would expect thermal emission to dominate, whereas for lower mass-loss rates inverse Compton would dominate. The mass-loss rate at which thermal emission begins to dominate depends on the value ${\epsilon}_r \, {\gamma}_{min}$. If ${\epsilon}_r << 1$ then we would expect that thermal emission would dominate at mass-loss rates  $\ge$ 10$^{-5}$ $\msun$ yr$^{-1}$. At the lower end of the RSG mass-loss rate function, $\approx10^{-7}$ $\msun$ yr$^{-1}$, the mass-loss rate is 2 orders of magnitude lower, and the thermal emission 4 orders of magnitude lower, whereas the non-thermal emission drops by only two orders of magnitude; therefore we would expect the non-thermal emission to dominate. In between these limits, as the mass-loss rate increases, we expect the contribution of thermal emission to increase and that of non-thermal emission to decrease. \\

 \item{\em Type IIn SNe:}  Type IIn SNe have the highest X-ray luminosities of all SN types. Furthermore, their emission has always been found to be thermal, as shown in \S~\ref{sec:spectra}. Therefore, the shock waves in Type IIn SNe must be expanding into a medium with very high mass-loss rates, generally higher than 10$^{-4} \;\msun$ yr$^{-1}$. The wind velocities quoted for Type IIn SNe progenitors are in the range of 50-150 km s$^{-1}$ (see review of Type IIn wind velocities in \cite{marcowithetal18}), which means that the mass-loss rate must be about a factor of 10 higher than what is shown on the plot. Thus many IIns will have mass-loss rates exceeding 10$^{-3}$ $\msun$ yr$^{-1}$.  Similar mass-loss rates have been derived by observations at other wavelengths such as X-ray, radio and optical. 
 
 A few years after explosion, IIns show a large deviation from the t$^{-1}$ slope for X-ray emission in a constant wind, and consequently from an r$^{-2}$ density profile for the wind medium, as seen in Figure~\ref{fig:lcmdot}. This suggests that the mass-loss rate, or wind velocity, or both, are functions of time, and hence radius. The slopes are much steeper than t$^{-1}$. If the wind velocity is assumed constant, as is frequently taken to be the case, then this implies that the mass-loss rate increases as one gets closer to the onset of core-collapse, and could exceed the rates suggested above. The mass-loss rate suggested for SN 2010jl at early times was $0.1\; \msun yr^{-1}$ \cite{franssonetal14}.
 
 Any model for the progenitors of Type IIn SNe must account for the decrease in mass-loss rate over time. A popular model for Type IIns is that their progenitors are LBV stars \cite{gl09}, which are thought to undergo {sudden eruptive mass-loss events}, providing the high mass-loss rates seen. It is not clear though how this would account for the decreasing mass-loss rates with time. Other possibilities for the high density may also exist. \cite{dwarkadas11} argued that clumpy winds, a prior LBV phase, or evolution in a wind-blown bubble followed by collision with a dense shell, as seen in SN 1987A, could account for the high mass-loss rates. The important point here is that the density is required to be high. It is only if this density results from a freely flowing wind that it may be equated to a mass-loss rate. If instead the density is due to a non-steady outflow, such as a surrounding disk, dense clumps in wind, or a dense shell of a wind bubble, then calculating a mass-loss rate from the density is meaningless. 
 
 In general it seems there may be many different progenitors of Type IIn SNe. Any system that can produce a high density ambient medium can lead to a Type IIn. In some cases, SNe of one type may encounter a high density somewhat later in their evolution, and transition to a Type IIn, as we discuss below.\\

 \item {\em Type IIL:}  Although the statistics are low, these have X-ray luminosities that are equivalent to or somewhat higher than Type IIP SNe. If their emission is thermal, it would mean they arise from progenitors with wind mass-loss rates comparable to, or somewhat larger than the IIPs. If mass-loss rates increase with stellar mass, this suggests progenitors for IILs with initial mass comparable to, or slightly exceeding that of the Type IIPs. It also appears as if there exists a continuum between IIP and IIL SNe. A caveat is that the statistics are quite low, especially for X-ray emission from SN IILs.\\
 
 \item {\em Type IIb:} The prototypical Type IIb SN, SN 1993J, shows thermal emission. Its mass-loss rate, as predicted by the plot, is somewhat higher than 10$^{-5}$ $\msun$ yr$^{-1}$. This is consistent with more accurate measurements that suggest a mass-loss rate of about 4 $\times 10^{-5}$ $\msun$ yr$^{-1}$ \cite{tatischeff09}. Most IIbs appear to have X-ray luminosities comparable or larger than the IIPs, thus suggesting mass-loss rates equivalent to or exceeding 10$^{-5}$ $\msun$ yr$^{-1}$. SN 2018gk \cite{boseetal21} has luminosities exceeding 10$^{41}$ erg s$^{-1}$, and therefore mass-loss rates exceeding 10$^{-4}$ $\msun$ yr$^{-1}$. \cite{boseetal21} estimate mass-loss rates as high as 2$\times 10^{-4}$ $\msun$ yr$^{-1}$ from the flash ionization features, consistent with our results. These could exceed the highest mass-loss rates known for RSGs, although they could be consistent with mass-loss rates for yellow hypergiants \cite{trungetal09,shenoyetal16}. This would assume that the wind velocity was $\approx$ 10 km s$^{-1}$.  If the velocities were higher, of order 1000 km s$^{-1}$, such as for compact W-R stars, then the mass-loss rates would be extremely high and this could be a case for a compact IIb. \cite{boseetal21}, however, suggest that the progenitor has a thin envelope of H, thus disallowing a W-R star. Their investigations suggest that the SN has a massive progenitor with initial mass between 19-26 $\msun$, also disfavoring a W-R star, {unless it was formed in a binary system}.
 
 It is possible that some of the other IIbs may have wind velocities that are more comparable to those of compact stars. In this case the mass-loss rates would be too high for RSGs, or even hypergiants, and may imply a non-thermal X-ray emission mechanism. Unfortunately observations are scarce, and there is no consitent method to measure the wind velocities. Based purely on the X-ray luminosity and spectra, we do not see evidence for compact progenitors of IIb SNe.\\

 \item { \em Type Ib/c:}  SNe of Type Ib/c are expected to arise from Wolf-Rayet (W-R) stars, due to the lack of H/He in their optical spectra. As pointed out, their X-ray spectra can be fit by both thermal and non-thermal models. Their position on the X-ray luminosity plot indicates a mass-loss rate $> 10^{-5}\; \msun$ yr$^{-1}$ if the X-ray emission is thermal. However, the winds of W-R stars are known to have velocities $> 1000$ km s$^{-1}$ \cite{crowther07}. This implies that the mass-loss rates quoted above must be multiplied by a factor $\ge$100. This would result in mass-loss rates for Type Ib/c progenitors $> 10^{-3} \;\msun$ yr$^{-1}$, substantially exceeding the known mass-loss rates of  W-R stars by about 2 orders of magnitude \cite{crowther07}. The inference is that the assumption of thermal emission is therefore incorrect. As mentioned earlier, \cite{cf06} have suggested that the emission from Ib/c SNe is non-thermal, due to inverse Compton processes at early times, and synchrotron emission at later times. That seems to be a much more plausible explanation for the high X-ray luminosities. 
 
There exists a class of Type Ib/c SNe that show increasing X-ray and/or radio emission at a delayed time after the SN explosion. This includes SNe 1996cr \cite{baueretal08,ddb10,quirolavasquezetal19}, 2001em \cite{cc06,chandraetal20}, 2004dk\cite{pooleyetal19}, and 2014c \cite{marguttietal17,thomasetal22,brethaueretal22}. These SNe essentially transitioned to a Type IIn SN months to years after outburst, as the SN shock encountered a higher density medium at late times. In each case, we would expect that the X-ray emission transitioned from non-thermal to thermal emission as the SN impacted the higher density. Unfortunately, this is difficult to demonstrate in practice. In 1996cr, 2001em, and 2014c, there are no initial high resolution X-ray observations, before the SN shock wave collided with the dense region. SN 2004dk did have a reasonably good XMM-Newton spectrum \cite{pooleyetal19}. Fitting the spectrum with a thermal model, the authors found that the temperature exceeded the range of XSPEC. A power-law model suggests a power-law fit with $\Gamma = 1.1 \pm 0.3$. Again it was not possible from the spectra to distinguish between the models, but a non-thermal fit implying synchrotron emission is certainly a possibility. At later times, as the X-ray emission was increasing, all the SNe showed signatures of thermal emission, as expected. The HETG spectrum of SN 1996cr was rich in X-ray emission lines \cite{ddb10,quirolavasquezetal19}, while the Chandra spectra of SN 2014c showed a clear Fe line. 

\cite{quirolavasquezetal19} suggested that there had to be an asymmetry in the SN ejecta, or surrounding medium, or both, in order to explain the extremely high resolution Chandra grating spectra of SN 1996cr. This asymmetry could be due to evolution in a binary system. \cite{pooleyetal19} suggested the existence of a wind bubble like region with a dense shell to explain the increasing X-ray emission in SN 2004dk. \cite{chandraetal20} attributed the increase in emission in SN 2001em to the interaction of the SN shock with a dense shell, ejected by the progenitor star due to common envelope evolution in a binary system. SN 2014c had the most accumulated data of all these SNe. \cite{thomasetal22} found a large discrepancy between the velocity evolution of various components, with the broad H$\alpha$ component having a FWHM of 2000 km s$^{-1}$, while the radio showed the shock expanding at $10,000 $ km s$^{-1}$. They explained the discrepancy in the various velocity components by suggesting a two-component surrounding medium, with a dense disk or torus in the equatorial region, and a lower density wind medium at higher latitudes. Such a medium could be formed by binary evolution, which has been suggested for the W-R progenitors of Type Ib/c SNe. \cite{brethaueretal22} echoed a similar explanation a few months later. \\

 \item { \em SLSNe:} Almost all observations of SLSNe-I are upper limits. There are two SLSNe-1 with actual detections. SCP06F6 was mentioned earlier. This is an outlier not only among SLSNe-I \cite{marguttietal18}, but in fact among all known SNe. The flux as mentioned before may be suspect, given the lack of a Chandra detection a few months later.
 
 X-ray emission was also noted in the slowly evolving SLSN-I PTF 12dam. However, \cite{marguttietal18} suggest that X-ray emission from the host galaxy may at least be in part responsible for the observed emission at the site of the SN. 
 
 A popular model for SLSNe-I is that they are powered by a central engine, such as a magnetar \cite{metzgeretal15}. The rotationally powered wind from the magnetar inflates a wind of electron/positron pairs that would heat the SN ejecta. The pairs could radiate X-rays or gamma-rays at the wind-termination shock via inverse Compton or sychrotron emission. However, except for the uncertain cases of SCP06F6 and PTF12dam, no X-ray or gamma-ray emission from a SLSN-I has been detected \cite{renaultetal18, andreonietal22, acharyyaetal23}
 
 SN 2006gy was the most optically luminous SNe ever recorded when it was detected. It would fall in the category of SLSNe-II. \cite{smithetal07} estimated the X-ray luminosity to be 1.65 $\times 10^{39}$ erg s$^{-1}$, which is quite low. Only 4 counts were detected in the extraction region, all below 2 keV, suggesting that the emission was all soft emission. For a wind-speed of 200 km s$^{-1}$, the authors calculated a mass-loss rate of 1.4 $\times 10^{4}$ M$_{\odot}$ yr$^{-1}$, given the observed luminosity. That is close to the mass-loss rate expected from Figure~\ref{fig:lcmdot}.  
 
 The unabsorbed luminosity for SN 2006gy was calculated using a column density to the source derived from the reddening, and did not include any additional column density due to a high-density circumstellar medium. This would raise the column density, and therefore the unabsorbed emission, considerably. The authors note that circumstellar interaction can not power the light curve, because the mass-loss rate needed for the that to happen is 3 orders of magnitude higher.  That, however, is a circular argument, because the mass-loss rate is computed from the luminosity assuming only the Galactic column density, i.e. under the assumption that \textit{there is no circumstellar medium around the SN}. So the low X-ray luminosity, and consequently the low mass-loss rate, may be underestimated.
 
While magnetars could power the hydrogen rich SLSNe-II, a more likely scenario is that they are powered by circumstellar interaction \cite{chatzopoulosetal13, moriyaetal13}.  This requires a large density in the surrounding medium, or alternatively a high mass CSM. It is then surprising that no X-ray emission is seen from the high density medium. It is possible that the medium is so optically thick that it absorbs all the X-rays, which are subsequently converted to optical or infrared emission. An alternative possibility is that the X-ray emission from SN 2006gy was underestimated as mentioned, due to underestimation of the column density.

\end{itemize}

In summary, non-thermal emission is characteristic of Type Ib/c SNe, and Type IIP SNe with low mass-loss rates. The other types of SNe generally show thermal emission, with the caveat that there are too few observations for Type IIL SNe. We do not see a division of Type IIb SNe into two different classes, but the low number statistics preclude a definitive answer.

Fast shocks in a low density medium, as expected for Type Ia SNe, Type Ib/c SNe, and the lower mass-loss rate Type IIP SNe, will result in non-thermal emission. Conversely, slower shocks in a higher density medium, such as in Type IIns, Type IIb, higher mass-loss rate Type IIPs, and the Type Ia-CSM SNe, will show thermal emission.

\subsection{Accurate determinations of mass-loss rates}
Using well-sampled X-ray lightcurves and high resolution X-ray data, it is possible to determine the mass-loss rates of individual SNe much more accurately. This was demonstrated in \cite{flc96}. From the X-ray lightcurve we can obtain $L_x$, and assuming $9 < n < 12$ \cite{mm99}, we can derive the value of `s'. Since the rate is a function of radius (or time),  the quantity $(\dot{M}/v_w)$ is computed at a specific reference radius (10$^{15}$ cm), from which it can be derived at any other radius.  Following \cite{flc96}, a spectral luminosity at 1 keV is used to calculate $(\dot{M}/v_w)$, although a similar expression can be derived at any other frequency. With well-sampled X-ray lightcurves and moderate resolution spectra, the mass loss rates of individual SNe can be better determined. \cite{chandraetal15} determined a mass-loss rate of 0.06 $\msun$ yr$^{-1} $ for SN 2010jl, which agrees well with the optically determined value of 0.1 $\msun$ yr$^{-1}$ by \cite{franssonetal14}. For SN 2005kd, \cite{drrb16} find a mass-loss rate of 4.3 $\times 10^{-4} \;\msun$ yr$^{-1}$ for a wind velocity of 10 km s$^{-1}$ at 10$^{16}$ cm, and infer that it could have been $> 10^{-3}\; \msun$ yr$^{-1}$ closer to core-collapse. If the wind velocity is higher, as is likely for Type IIn SNe, the mass-loss rate would be higher.

\section{Discussion and Conclusions:} We have accumulated, and presented herein, the X-ray data on most young SNe that have been observed to date. We find that over 115 SNe have been observed in X-rays, although only about {60\%} of them have been detected. As a class the IIn SNe are the most luminous, whereas the IIPs have the lowest luminosity.  The IIbs, IILs, and Ib/c SNe all seem to lie in between these limits. Most SNe have X-ray luminosities between 10$^{36}$ and 10$^{42}$ erg s$^{-1}$. At the lower luminosity end we have only one SN, SN 1987A, which happens to be the closest SNe to us discovered in the modern era. There is presumably a large population of low luminosity SNe, many of them evolving in wind blown bubbles similar to SN 1987A, that we are missing.

We studied the spectra of representative SNe within each class, in order to determine the emission mechanism, which could be inverse Compton, synchrotron or thermal bremsstrahlung. It turns out that this is difficult to determine only from the spectra alone. While some well studied SNe may have thousands of counts in the spectrum, most have only a few tens to hundreds of counts. In many cases, spectra can be equally well fitted by both thermal and nonthermal models along with absorption.  Type IIns, which have the highest luminosities, clearly show thermal spectra with well defined lines. These SNe should be interacting with a high density medium. Their X-ray luminosity falls steeply after a few thousand days. Type Ib/c SNe must have non-thermal emission, either inverse Compton or synchrotron. Their X-ray luminosities are too high to be attributable to thermal emission. This is due to the large wind velocity of the progenitor stars, assumed to be W-R stars, which results in a low density in the surrounding medium, into which the SN shock expands. This low density cannot account for the high X-Ray luminosity.

Type IIb SNe such as 1993J clearly appear to be thermal. From the X-ray standpoint alone, there is no clear division into two classes of IIb SNe. Type IIPs all appear to have mass-loss rates below about 10$^{-5}\, \msun$ yr$^{-1}$. It is likely that those with mass-loss rate on the order of 10$^{-5} \,\msun$ yr$^{-1}$ may have thermal emission, while for those with mass-loss rates two orders of magnitude lower, non-thermal emission will dominate.

The X-ray emission from young SNe, {if thermal in origin}, can be used to deduce the density of the medium into which the SN is expanding, and thus the mass-loss rate of the progenitor star for a  medium formed by stellar mass-loss. This can provide insight into the mass-loss rates of massive stars in the years, decades and centuries before core-collapse, which are otherwise  difficult to determine. The values so derived agree approximately with those obtained from observations at other wavelengths. They show that Type IIPs are expanding in winds with the lowest mass-loss rates, while IIns have the highest mass-loss rates, which may exceed 10$^{-3} \;\msun$ yr$^{-1}$, and perhaps higher depending on their wind velocities. It is unclear which progenitor stars could sustain such high mass-loss rates. {If the emission is non-thermal, it can still be used to determine the mass-loss rate (see equation~\ref{eq:ic}) but there are more parameters to contend with, and they may not be well determined.}

It should be emphasized here that while we frequently talk of a single parameter, the mass-loss rate, the important parameter is the density of the medium. For a freely flowing wind this is dependent on $\dot{M}/v_w$. For simplicity we normally assume the velocity to be constant, so the terms mass-loss rate and density are used interchangably. But it is important to keep in mind that the density may not always be due to a free-flowing wind. {Some of the very high mass-loss rates attributed to Type IIn and other SNe may fall in this category.  Indeed, for many IIns, such as SN 1986J, SN1996cr, SN 2004dk, and SN 2014c, the high density has been attributed to clumps in the wind \cite{chugai93,cd94}, a dense shell formed by a wind-blown bubble \cite{dwarkadas11,pooleyetal19} or a disk or torus formed by binary evolution \cite{quirolavasquezetal19,chandraetal20,thomasetal22}.  In this scenario the concept of mass-loss rate is not meaningful, as the density is not due to a steady outflow. }

\vspace{6pt} 





\authorcontributions{N/A}

\funding{This research received no external funding}

\institutionalreview{N/A}

\informedconsent{N/A}

\dataavailability{The sources of all data are cited in Table~\ref{tab:xraysne}.  }

\acknowledgments{This research has made use of data obtained from the Chandra Data Archive provided by the Chandra X-ray Center (CXC). The scientific results reported in this article are based to a significant degree on observations made by the Chandra X-ray Observatory, including ObsIDs 794, 4417, 735, 3870, 6727 and 4631. We have also used XMM-Newton observation 0410581101. This research has made use of software provided by the Chandra X-ray Center (CXC) in the application packages CIAO and Sherpa. It has also made use of the Chandra PIMMS and Colden packages. We have also used the XMM-SAS software package for XMM data. {I thank all collaborators who have worked with me over the years on exploring X-ray SNe. My X-ray SNe research has been supported by granting agencies including NASA ADAP, NSF, and Chandra, to whom I am grateful. I thank the referees for a detailed reading of the paper, and for their comments and suggestions.}}

\conflictsofinterest{ The author declares no conflicts of interest.} 




\reftitle{References}


\bibliography{ref}

\PublishersNote{}
\end{document}